\documentclass[journal]{IEEEtran}
\usepackage{cite}
\usepackage{amsmath,amssymb,amsfonts}
\usepackage{algorithmic}
\usepackage{graphicx}
\usepackage{grffile}
\usepackage{textcomp}
\usepackage{epstopdf}
\usepackage{epsfig}
\usepackage{subfigure}
\usepackage{psfrag}
\usepackage{color,soul}
\usepackage{euscript}

\def\BibTeX{{\rm B\kern-.05em{\sc i\kern-.025em b}\kern-.08em
    T\kern-.1667em\lower.7ex\hbox{E}\kern-.125emX}}

\newcommand{\dya}[1]
{\overline{\overline{#1}}}

\newcommand{\dm}[1]
{\dya{\mathcal{#1}}}

\newcommand{\vm}[1]
{\overline{\mathcal{#1}}}

\begin{document}
\title{Wave Scattering by a Cylindrical Metasurface Cavity of Arbitrary Cross-Section: \\ Theory and Applications}
\author{Mojtaba Dehmollaian, \IEEEmembership{Senior Member, IEEE}, Nima Chamanara, and Christophe Caloz, \IEEEmembership{Fellow, IEEE}
\thanks{The paper is submitted on 27 August 2018. \newline
M. Dehmollaian is with the Center of Excellence on Applied Electromagnetic Systems, School of Electrical and Computer Engineering, University of Tehran, 15719-14911, Tehran, Iran (e-mail: m.dehmollaian@ut.ac.ir). \newline
N. Chamanara and C. Caloz are with the Department of Electrical Engineering, Polytechnique Montréal, H3T 1J4, Québec, Canada (e-mail: nimachamanara@gmail.com and christophe.caloz@polymtl.ca).}}
\maketitle

\begin{abstract}
This paper presents a technique, combining the integral equations (IE) and the Generalized Sheet Transition Conditions (GSTCs) with bianisotropic susceptibility tensors, to compute electromagnetic wave scattering by cylindrical metasurfaces -- forming two-dimensional porous cavities -- of arbitrary cross sections. Moreover, it applies this technique to two problems -- cloaking with circular and rhombic shapes and illusion optics with an elliptic shape -- that both validate it, from comparison with specifications used in an exact synthesis of the metasurfaces, and reveal interesting capabilities of such metasurface structures. Particularly, active cylindrical metasurfaces can perfectly cloak and hence eliminate the extinction cross section of various cylindrical shapes, and simple purely passive versions of them, practically more accessible, still perform quite good cloaking and provide remakable extinction cross section reduction.
\end{abstract}


%\tableofcontents

\section{Introduction}
\label{sec:introduction}

\IEEEPARstart{A}{Metasurface} is a two-dimensional (2D) array of subwavelength scattering particles that transforms electromagnetic waves in various fashions~\cite{HKS}. They have been used in a myriad of applications, including wave-front transformers~\cite{Gr}, waveguide walls~\cite{HKN}, leaky-wave antennas~\cite{Maci}, reflector antennas~\cite{AM}, remote controllers~\cite{ALSC}, illusion devices~\cite{Abd}, computational imagers~\cite{IS}, light-extraction enhancement cavities~\cite{CAKC}, nanoparticle optical force shapers~\cite{Grb} and solar sails~\cite{AC17}. A recent overview of the concept, design and applications of metasurfaces may be found for instance in~\cite{AC}.

A metasurface is essentially a \emph{sheet discontinuity} of space for electromagnetic fields~\cite{Id}. It may be effectively modeled by \emph{bianisotropic surface susceptibility tensors}, which relate the difference and average fields on either side of it via \emph{electric and magnetic surface current polarization densities} from the Huygens principle. This leads to generalized boundary conditions called the Generalized Sheet Transition Conditions~(GSTCs)~\cite{ricoy}. GSTCs are extremely powerful because 1)~they provide deep insight into the physics of metasurface scattering, 2)~they offer dramatic computational time saving by replacing the physical slab metasurface structure, that would require very dense meshing, by a much simpler sheet structure, without compromising accuracy given the subwavelength thickness of the metasurface, 3)~they represent a fundamental generalization of conventional boundary conditions, as explicitly shown in~\cite{JVCY}.

Once a metasurface has been \emph{synthesized according to specifications}~\cite{ASC}, it must be \emph{numerically analyzed} both for validating the specification and for determining the response to parameter values (incidence angle, polarization, frequency, etc.) different from the specification, for full characterization. Much effort has been recently dedicated to such numerical analysis~\cite{VC}, using the main computational techniques coupled with GSTCs. This includes finite-difference frequency domain (FDFD)~\cite{VAC}, finite-difference time-domain (FDTD)~\cite{SG,VCC}, the finite-element method (FEM)~\cite{SJC} and the spectral-domain integral equation (IE)~\cite{CAC}.

The computational studies mentioned in the previous paragraph have been so far restricted to \emph{planar} metasurfaces. However, practical applications involve a diversity of \emph{curved} objects, whose metasurface enhancement would naturally require curved surfaces. For example, scattering radar cross section (RCS) reduction structures, conformal antenna radomes, and cloaking (e.g.~\cite{Moreno}) and illusion devices have strongly curved geometries. There is therefore a pressing need for extending current computational techniques for planar metasurfaces to curved metasurfaces. Early efforts in this direction include the FDTD analysis of spherical metasurface cavities~\cite{JVC} and arbitrarily curved metasurfaces~\cite{Genevet}, and the method of moments (MoM) analysis of circular-cylindrical metasurfaces~\cite{SH}.

This paper presents a generalization of the IE-MoM analysis technique of the circular-section cylindrical metasurface using circular cylindrical coordinates in~\cite{SH} to a cylindrical metasurface with arbitrary cross section. Moreover, using this technique, it demonstrates that a properly synthesized metasurface, coating circular cross section and rhombic cross section cylinders, leads to perfect active cloaking and extinction cross width reduction while taking a naive purely passive version of the metasurfaces still provide excellent cloaking and substantial extinction cross width, and also demonstrates the optical illusion capability of such metasurfaces in the case of an elliptical structure.

The paper is organized as follows. Section~\ref{sec:desc} describes the problem, outlines the Huygens approach used next, and lays out some mathematical preliminaries. Section~\label{sec:formul} presents the GSTC-IE formulation, culminating with a simple and insightful matrix system solution, and numerically validates the formulation by comparison with analytical results for a circular cross-section metasurface. Section~\ref{sec:sim} presents cloaking (external source) and illusion (internal source) application examples, which further confirm the method, and demonstrates practically remarkable properties of circular, rhombic and elliptical metasurface cavities. Finally, Section~\ref{sec:con} provides concluding remarks.

\begin{figure}[!t]
\centerline{\includegraphics[width=3 in, height = 2.4 in]{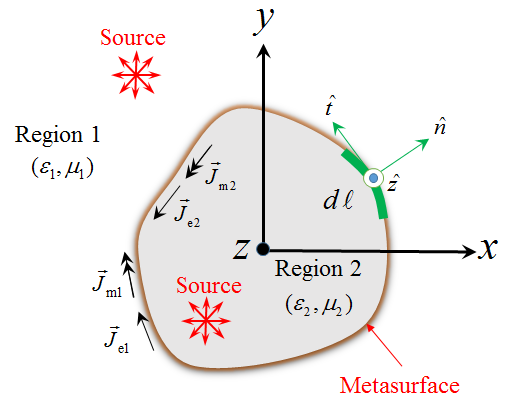}}
\caption{Problem to be solved: analysis of wave scattering by a metasurface cavity of arbitrary cross section, under illumination by an arbitrary source placed either outside or inside of it, using equivalent electric and magnetic polarization currents. Given its metasurface enclosure, the cavity is porous in nature, and may be de facto be even open, at locations of zero currents.\label{prob}}
\end{figure}

\section{Problem Description, Huygens Approach and Mathematical Preliminaries}\label{sec:desc}

Figure \ref{prob} shows the problem to be solved, namely the analysis of wave scattering by a metasurface cavity of arbitrary cross section illuminated by an arbitrary source distribution. This is a two-dimensional problem, with $\partial/\partial z\equiv 0$, and line source Green function. The exterior region, denoted by the subscripts~1, consists of a medium with permittivity $\varepsilon_1$ and permeability $\mu_1$, while the interior region, denoted by the subscripts~2, consists of a medium with permittivity $\varepsilon_2$ and permeability $\mu_2$. Given its metasurface enclosure, the cavity is fundamentally \emph{porous}, i.e. allowing partial field transmission through it, and may even accommodate complete \emph{openings} at parts of its contour.

We will next need to use the global (Cartesian) coordinate system $(x,y,z)$ \emph{and} the coordinate system local to the metasurface $(n,t,z)$, with $\hat t = \hat z \times \hat n$ , shown in Fig.~\ref{prob}. Expressing the metasurface contour by the polar function $\rho=f(\phi)$ yields the following relations between the local normal unit vector and the global unit vectors $\hat{\rho}$ and $\hat{\phi}$: $\hat n = [\hat \rho  - (f'/f)\hat \phi ]/\sqrt {1 + {{(f'/f)}^2}}$.

The problem will be solved in terms of the \emph{Huygens principle}, where the metasurface enclosure is replaced by equivalent exterior and interior electric and magnetic polarization currents, $\vec {J}_\text{e1}$, $\vec {J}_\text{m1}$, $\vec {J}_\text{e2}$, and $\vec {J}_\text{m2}$, respectively, ``radiating'' the scattered fields everywhere in space. This will be accomplished in the following three successive steps. First, we will apply the GSTCs in the \emph{local} coordinate system of the metasurface, $(n,t,z)$, to relate the exterior and interior tangential electric and magnetic fields, $\vec E_{\| 1}$, $\vec H_{\| 1}$, $\vec E_{\| 2}$ and $\vec H_{\| 2}$, respectively, related to the polarization currents by
\begin{subequations}\label{eq:curr_tang_fields}
\begin{equation}
\vec {J}_\text{e1}=\hat n \times {{\vec H}_{\| 1}},\quad
\vec {J}_\text{m1}=-\hat n \times {{\vec E}_{\| 1}},
\end{equation}
\begin{equation}
\vec {J}_\text{e2}=-\hat n \times {{\vec H}_{\| 2}},\quad
\vec {J}_\text{m2}=\hat n \times {{\vec E}_{\| 2}}
\end{equation}
\end{subequations}
\noindent to the bianisotropic surface susceptibility tensors characterizing the metasurface. This will result in a pair of equations. Second, we will apply the Huygens principle formula to express the fields produced by these tangential fields at the surface of the metasurface in terms of electric field IEs, first in the \emph{global} coordinate system, $(x,y,z)$, where analytical Green functions are available, and next, using proper coordinate transformation, in the local coordinate system, for compatibility with the GSTC equations. This will result in another pair of equations. Third, we will combine the pairs of equations obtained in the first (GSTCs) and second (IEs) steps, which will lead to a system of four equations in four unknowns, the tangential fields $\vec H_{\| 1}$, $\vec E_{\| 1}$, $\vec H_{\| 2}$ and $\vec E_{\| 2}$ in~\eqref{eq:curr_tang_fields}, that will be solved by MoM. Once these unknowns have been determined, they will substituted into~\eqref{eq:curr_tang_fields}, and the resulting expressions will be further inserted into the IE, which will provide the fields everywhere in space.

To write the GSTCs in the metasurface-local coordinate systems, we will need to express all the field and current vectors in this system. For instance,
\begin{equation}
\vec{E}_{\|1}=
\begin{pmatrix}
  E_{1t} \\
  E_{1z}
\end{pmatrix},
\end{equation}
and similar expressions hold for
$\vec{E}_{\|2}$, $\vec{H}_{\|1,2}$, $\vec{J}_{\text{e}1,2}$, $\vec{J}_{\text{m}1,2}$. Moreover, the polarization currents~\eqref{eq:curr_tang_fields} in this coordinate system may be conveniently expressed in terms of the corresponding tangential fields as
\begin{subequations}\label{eq:Nte_def}
\begin{equation}\label{eq:Je_vs_N}
\vec {J}_\text{e1}
= {H_{1t}}\hat z - {H_{1z}}\hat t
=\begin{pmatrix}
-H_{1z} \\ H_{1t}
\end{pmatrix}
=\overline{\overline {\text{N}}}\cdot\vec H_{\| 1},
\end{equation}
with
\begin{equation}\label{eq:Ntens}
\overline{\overline {\text{N}}}
=\begin{pmatrix}
0&-1\\
1&0
\end{pmatrix},
\end{equation}
and
\begin{equation}
{{\vec J}_\text{m1}}=-\dya{\text{N}}\cdot\vec E_{\| 1},\quad
{{\vec J}_\text{e2}}=-\dya{\text{N}}\cdot\vec H_{\| 2},\quad
{{\vec J}_\text{m2}}=\dya{\text{N}}\cdot\vec E_{\| 2}.
\end{equation}
\end{subequations}

As mentioned above, the application of the IEs will first require expressing all the metasurface-tangential quantities in the global coordinate system. This may be accomplished using the local-to-global $3\times 2$ tangential transformation tensor
\begin{equation}\label{eq:T_tensor}
\overline{\overline{T}}(\vec \rho)=
\begin{pmatrix}
{\hat x \cdot \hat t}&0\\
{\hat y \cdot \hat t}&0\\
0&1
\end{pmatrix},
\end{equation}
which projects the $t$-components of a right-operated vector onto the $x$ and $y$ directions of the global system, while leaving the shared coordinate $z$ unchanged. We have then, for instance,
\begin{equation}\label{tr}
\begin{pmatrix}
\mathcal{J}_{\text{e}x}(\vec{\rho}) \\
\mathcal{J}_{\text{e}y}(\vec{\rho}) \\
\mathcal{J}_{\text{e}z}(\vec{\rho})
\end{pmatrix}
={{\vec {\mathcal{J}}}_\text{e1}}(\vec{\rho})
=\overline{\overline{T}}(\vec \rho)\cdot
\begin{pmatrix}
J_{\text{e}1t} \\
J_{\text{e}1z}
\end{pmatrix}
=\overline{\overline{T}}(\vec \rho)\cdot{{\vec J}_\text{e1}},
\end{equation}
where the local-system current ${{\vec J}_\text{e1}}$ transforms into the global-system current ${{\vec {\mathcal{J}}}_\text{e1}}(\vec{\rho})$ with $\vec{\rho}=x\hat{x}+y\hat{y}$. Inserting~\eqref{eq:Je_vs_N} into~\eqref{tr} further yields
\begin{subequations}\label{gc}
\begin{equation}
\vec{\mathcal{J}}_\text{e1}(\vec{\rho}) = \dya{T}(\vec \rho) \cdot \dya{\text{N}} \cdot \vec H_{\| 1},
\end{equation}
and similarly
\begin{equation}
\vec{\mathcal{J}}_\text{m1}(\vec{\rho}) = -\dya{T} \cdot \dya{\text{N}} \cdot \vec E_{\| 1}
\end{equation}
and
\begin{equation}
\vec{\mathcal{J}}_\text{e2}(\vec{\rho}) = -\dya{T} \cdot \dya{\text{N}} \cdot \vec H_{\| 2},\quad
\vec{\mathcal{J}}_\text{m2}(\vec{\rho}) = \dya{T} \cdot \dya{\text{N}} \cdot \vec E_{\| 2},
\end{equation}
\end{subequations}
which conveniently express the global-system currents in terms of the corresponding local-system tangential fields.

Finally, in this application of the MoM, we will need to have expressions for the tangential fields in the local coordinate system in terms of the fields in the global coordinate system. Such expressions may be obtained by pre-multiplying the latter by the transpose of the tensor $\overline{\overline{T}}(\vec \rho)$ in~\eqref{eq:T_tensor}. Indeed, for instance, the tangential electric field in region~1 expressed in the local coordinate system, ${\vec E}_{\| 1}$, may be written as
\begin{equation}\label{gtl}
\vec E_{\| 1}=
\begin{pmatrix}
  E_{1t} \\
  E_{1z}
\end{pmatrix}
=
\begin{pmatrix}
{\hat x \cdot \hat t}&{\hat y \cdot \hat t}&0\\
0&0&1
\end{pmatrix}
\begin{pmatrix}
  \mathcal{E}_{1x} \\
  \mathcal{E}_{1y}
\end{pmatrix}
=
{\overline {\overline{T}}^\text{t}}(\vec \rho) \cdot \vec{\mathcal{E}}_1(\vec \rho)
\end{equation}
where superscript $\text{t}$ denotes the transpose and $\vec{\mathcal{E}}_1(\vec{\rho})$ is the electric field in region~1 expressed in the global coordinate system.

In the remainder of the paper, the variables $c$, $\omega$, $k$, $\eta$, ${\epsilon_0}$, and ${\mu_0}$ respectively represent the speed of light, the angular frequency, the wavenumber, the wave impedance, and the permittivity and permeability of free-space, and the time convention $e^{-i \omega t}$ is implicitly assumed.

\section{Computational Formulation}\label{sec:formul}

\subsection{Generalized Sheet Transition Conditions (GSTCs)}

Let us assume, for simplicity, that the metasurface has only \emph{metasurface-tangential} (or, equivalently, \emph{transverse}) electric and magnetic surface polarization densities, ${\vec P}_\|$ and ${\vec M}_\|$, respectively, and zero normal components, i.e. $P_\perp=M_\perp=0$ or
$P_n=M_n=0$. In this case, the harmonic ($e^{-i\omega t}$) GSTCs reduce to~\cite{AC}
\begin{subequations}\label{GSTC}
\begin{equation}\label{eq:GSTC_DE}
\hat n \times \Delta \vec E = i\omega {\mu _0}{{\vec M}_\|},
\end{equation}
\begin{equation}
\hat n \times \Delta \vec H =  -i\omega {{\vec P}_\|},
\end{equation}
\end{subequations}
where $\Delta$ denotes the difference of the fields at both sides of the metasurface, and where
\begin{subequations}\label{MP}
\begin{align}\label{MPa}
{{\vec M}_\|} = \left( \begin{array}{*{20}{c}}
{{M_t}}\\
{{M_z}}
\end{array} \right) = & \frac{1}{Z_0}\left( \begin{array}{*{20}{c}}
{\chi _\text{me}^{tt}}&{\chi _\text{me}^{tz}}\\
{\chi _\text{me}^{zt}}&{\chi _\text{me}^{zz}}
\end{array} \right)\left( \begin{array}{*{20}{c}}
{E_{t,\,\text{av}}}\\
{E_{z,\,\text{av}}}
\end{array} \right) \nonumber \\
& + \left( \begin{array}{*{20}{c}}
{\chi _\text{mm}^{tt}}&{\chi _\text{mm}^{tz}}\\
{\chi _\text{mm}^{zt}}&{\chi _\text{mm}^{zz}}
\end{array} \right)\left( \begin{array}{*{20}{c}}
{H_{t,\,\text{av}}}\\
{H_{z,\,\text{av}}}
\end{array} \right) \nonumber \\ & ={\frac{1}{\eta_0}} \overline {\overline {\chi}} _{\|,\,\text{me}} \cdot \vec E_{\|,\,\text{av}} + \overline{\overline {\chi}} _{\|,\,\text{mm}} \cdot \vec H_{\|,\,\text{av}},
\end{align}
\begin{equation}
{{\vec P}_\|} = \left( \begin{array}{*{20}{c}}
{{P_t}}\\
{{P_z}}
\end{array} \right) = {\varepsilon _0} \overline {\overline {\chi}} _{\|,\,\text{ee}} \cdot \vec E_{\|,\,\text{av}} + \frac{1}{c} \overline{\overline {\chi}} _{\|,\,\text{em}} \cdot \vec H_{\|,\,\text{av}},
\end{equation}
\end{subequations}
where $\overline{ \overline {\chi}}_\text{mm}$, $\overline{ \overline {\chi}}_\text{me}$, $\overline{ \overline {\chi}}_\text{em}$ and $\overline{ \overline {\chi}}_\text{ee}$ are the magnetic-to-magnetic, electric-to-magnetic, matnetic-to-electric and electric-to-electric surface susceptibilitiy tensors, respectively, and where ``av'' denotes the average of the fields at both sides of the metasurface~\cite{AC}.

Inserting (\ref{MP}) into (\ref{GSTC}) and using~\eqref{eq:Ntens} conveniently expresses the GSTCs in terms of the tangential fields, namely
\begin{subequations}\label{GSTC2}
\begin{align}
& \overline {\overline {\text{N}}} \cdot (\vec E_{\|1} - \vec E_{\|2}) = \nonumber \\
& \frac{{i\omega {\mu _0}}}{{2{Z_0}}}\overline{ \overline {\chi}}_{\|,\text{\,me}} \cdot (\vec E_{\|1}+ \vec E_{\|2}) + \frac{{i\omega {\mu _0}}}{2}\overline{ \overline {\chi}} _{\|,\text{\,mm}} \cdot (\vec H_{\|1}+ \vec H_{\|2}),
\end{align}
\begin{align}
& \overline {\overline{\text{N}}} \cdot (\vec H_{\|1} - \vec H_{\|2}) =  \nonumber \\
-&\frac{{i\omega {\varepsilon _0}}}{2}\overline {\overline {\chi}}_{\|,\text{\,ee}} \cdot (\vec E_{\|1} + \vec E_{\|2}) - \frac{{i\omega }}{{2c}}\overline{\overline {\chi}}  _{\|,\text{\,em}} \cdot (\vec H_{\|1}+ \vec H_{\|2}),
\end{align}
\end{subequations}
which provide two equations in terms of ${\vec E_{\|1}}$, ${\vec H_{\|1}}$, ${\vec E_{\|2}}$ and ${\vec H_{\|2}}$, respectively, in the local coordinate system.

\subsection{Integral Equation (IE)}

According to the Huygens principle, the fields scattered by the metasurface outside and inside the metasurface enclosure may be expressed as the convolution of the equivalent electric and magnetic polarization currents with the dyadic Green function of the corresponding medium $\overline{\overline {\mathcal{G}}}_{1,2}(\vec{\rho},\vec{\rho}')$ as~\cite{Tsang} (Eq.~(2.2.11))
\begin{subequations}\label{eq:Hp_IE}
\begin{align}\label{Eq1}
\vm{E}_1(\vec \rho ) - \vm{E}_\text{i}(\vec \rho )
=&
i{k_1}{\eta_1}\oint {\dm{G}_1 (\vec \rho ,\vec \rho ') \cdot \vm{J}_\text{e1}(\vec \rho ')\,d\ell '} \nonumber \\
& - \oint {\nabla  \times \dm{G}_\text{1} (\vec \rho ,\vec \rho ') \cdot \vm{J}_\text{m1}(\vec \rho ')\,d\ell '},
\end{align}
\begin{align}\label{Eq2}
\vm{E}_2(\vec \rho)
=&
i{k_2}{\eta_2}\oint {\dm{G}_2 (\vec \rho ,\vec \rho ') \cdot \vm{J}_\text{e2}(\vec \rho ')\,d\ell '} \nonumber \\
& - \oint {\nabla  \times \dm{G}_\text{2} (\vec \rho ,\vec \rho ') \cdot \vm{J}_\text{m2}(\vec \rho ')\,d\ell '},
\end{align}
\end{subequations}
where $\vm{E}_1$ and $\vm{E}_2$ are the total electric fields in regions 1 and 2, respectively, and  $\vm{E}_{\text{i}}$ is the incident field, all in the global coordinate system. Here, the incident field is illuminating the structure from the outside, i.e. region 1, of the cavity; the inside excitation case will be treated below. The line integrals run along the cylindrical boundary of the metasurface (Fig. \ref{prob}). The two-dimensional dyadic Green function and its curl are given by \cite{Tsang} (pp. 54-55) with $\partial/\partial z \equiv 0 $ as
\begin{subequations}
\begin{equation}\label{G1}
\dm{G}(\vec \rho ,\vec \rho ')
=\begin{pmatrix}
{1 + \frac{1}{{{k^2}}}\frac{{{\partial ^2}}}{{\partial {x^2}}}}&{\frac{1}{{{k^2}}}\frac{{{\partial ^2}}}{{\partial x\partial y}}}&0\\
{\frac{1}{{{k^2}}}\frac{{{\partial ^2}}}{{\partial y\partial x}}}&{1 + \frac{1}{{{k^2}}}\frac{{{\partial ^2}}}{{\partial {y^2}}}}&0\\
0&0&1
\end{pmatrix}g(\vec{\rho},\vec{\rho}')
\end{equation}
and
\begin{equation}\label{F1}
\dm {F}(\vec \rho ,\vec \rho ') =\nabla \times \dm{G} (\vec \rho ,\vec \rho ') =
\begin{pmatrix}
0&0&{\frac{\partial }{{\partial y}}}\\
0&0&{-\frac{\partial }{{\partial x}}}\\
{-\frac{\partial }{{\partial y}}}&{\frac{\partial }{{\partial x}}}&0
\end{pmatrix}g(\vec{\rho},\vec{\rho}'),
\end{equation}
where
\begin{equation}\label{2DG}
g(\vec{\rho},\vec{\rho}')={i}\,H_0^{(1)}(k|\vec{\rho} - \vec{\rho} '|)/\text{4}
\end{equation}
\end{subequations}
is the two-dimensional scalar Green function, with $H_0^{(1)}(\cdot)$ being the first-kind Hankel function of zeroth order, and  $\vec \rho  = (x,y)$ and $\vec \rho ' = (x',y')$ are the observation and source points, respectively. Defining $\Delta x = x - x'$, $\Delta y = y - y'$, and hence $\Delta \rho  = |\vec \rho  - \vec \rho '|=\sqrt{\Delta x ^2 + \Delta y ^2}$ allows one to eliminate the spatial derivatives in~\eqref{G1} and \eqref{F1}, which simplify then to
\begin{align}\label{G}
& \dm{G} = \nonumber \\
& \left( \begin{array}{*{20}{c}}
{{{(\Delta y)}^2}}&{\frac{{\Delta x\,\Delta y}}{{k\,}}\,}&0\\
{\frac{{\Delta x\,\Delta y}}{{k\,}}}&{{{(\Delta x)}^2}}&0\\
0&0&{{{(\Delta \rho )}^2}}
\end{array} \right)\frac{i}{{{{4(\Delta \rho )}^2}}}H_0^{(1)}(k\Delta \rho ) \nonumber \\
+ &\left( \begin{array}{*{20}{c}}
h&{ - 2\frac{{\Delta x\,\Delta y}}{k}}&0\\
{ - 2\frac{{\Delta x\,\Delta y}}{k}}&-h&0\\
0&0&0
\end{array} \right)
\frac{i}{{4k{{(\Delta \rho )}^3}}}H_1^{(1)}(k\Delta \rho )
\end{align}
and
\begin{equation}
\dm{F} = \left( \begin{array}{*{20}{c}}
0&0&{\Delta y}\\
0&0&{ - \Delta x}\\
{ - \Delta y}&{\Delta x}&0
\end{array} \right)\,\,\frac{{ - ik}}{{4\Delta \rho }}H_1^{(1)}(k\Delta \rho ),
\label{F}
\end{equation}
where $h={{{(\Delta x)}^2} - {{(\Delta y)}^2}}$ and $H_1^{(1)}(\cdot)$ is the first-kind Hankel function of first order.

Next, Eqs.~\eqref{eq:Hp_IE} are solved using MoM with the point matching technique. For this purpose, the metasurface boundary is discretized into $N$ segments with lengths $\Delta {\ell _n}$, $n=1,2,\ldots,N$. For the $m^{\text{th}}$ segment, we have
\begin{subequations}\label{MM}
\begin{align}
& {{\vm E}_1}({{\vec \rho }_m}) -  {{\vm E}_{\text{i}}}({{\vec \rho }_m}) = \nonumber \\ &i{k_1}{\eta_1} \sum\limits_{n = 1}^N {\dm{G}_1({{\vec \rho }_m},{{\vec \rho }_n}) \cdot {{\vm J}_\text{e1}}({{\vec \rho }_n})\,\Delta {\ell _n}} \nonumber \\
& -  \sum\limits_{n = 1}^N {\dm{F}_1({{\vec \rho }_m},{{\vec \rho }_n}) \cdot {{\vm J}_\text{m1}}({{\vec \rho }_n})\,\Delta {\ell _n}},
\end{align}
and
\begin{align}
& {{\vm E}_2}({{\vec \rho }_m}) = \nonumber \\
&i{k_2}{\eta_2} \sum\limits_{n = 1}^N {\dm{G}_2({{\vec \rho }_m},{{\vec \rho }_n}) \cdot {{\vm J}_\text{e2}}({{\vec \rho }_n})\,\Delta {\ell _n}} \nonumber \\
&- \sum\limits_{n = 1}^N {\dm{F}_2({{\vec \rho }_m},{{\vec \rho }_n}) \cdot {{\vm J}_\text{m2}}({{\vec \rho }_n})\,\Delta {\ell _n}}.
\end{align}
\end{subequations}
Multiplying both sides by $\dya{T}^\text{t}$ [Eq.~\eqref{eq:T_tensor}] as in \eqref{gtl} and expressing the surface currents in terms of the tangential fields $\vec H_{\| 1}$, $\vec E_{\| 1}$, $\vec H_{\| 2}$, and $\vec E_{\| 2}$ using (\ref{gc}) yields
\begin{subequations}\label{MM12}
\begin{align}
&\vec E_{\| 1}({{\vec \rho }_m}) - \vec E_{\| \text{i}}({{\vec \rho }_m}) = \nonumber \\ &i{k_1}{\eta_1}\dya{T}^{\text{t}}({\vec \rho }_m) \cdot \sum\limits_{n = 1}^N {\dm{G}_1({{\vec \rho }_m},{{\vec \rho }_n}) \cdot \left[\dya{T}(\vec \rho_n) \cdot \dya{\text{N}}\cdot\vec H_{\| 1}(\vec \rho_n)\right]\Delta {\ell _n}} \nonumber \\
& + \dya{T}^{\text{t}}({{\vec \rho }_m}) \cdot \sum\limits_{n = 1}^N {\dm{F}_1({{\vec \rho }_m},{{\vec \rho }_n}) \cdot \left[\dya{T}(\vec \rho_n) \cdot \dya{\text{N}}\cdot\vec E_{\| 1}(\vec \rho_n)\right]\Delta {\ell _n}},
\end{align}
\begin{align}
&\vec E_{\| 2}({{\vec \rho }_m})=\nonumber \\ &-i{k_2}{\eta_2}\dya{T}^{\text{t}}({\vec \rho }_m) \cdot \sum\limits_{n = 1}^N {\dm{G}_2({{\vec \rho }_m},{{\vec \rho }_n}) \cdot \left[\dya{T}(\vec \rho_n) \cdot \dya{\text{N}}\cdot\vec H_{\| 2}(\vec \rho_n)\right]\Delta {\ell _n}}\nonumber \\
&-\dya{T}^{\text{t}}({{\vec \rho }_m}) \cdot \sum\limits_{n = 1}^N {\dm{F}_2({{\vec \rho }_m},{{\vec \rho }_n}) \cdot \left[\dya{T}(\vec \rho_n) \cdot \dya{\text{N}}\cdot\vec E_{\| 2}(\vec \rho_n)\right]\Delta {\ell _n}},
\end{align}
\end{subequations}
where $\vec E_{\| \text{i}}({{\vec \rho }_m})=\dya{T}^{\text{t}}\cdot\vm E_{\text{i}}({{\vec \rho }_m})$ is the incident tangential electric field in the local coordinate system. Equations~\eqref{MM12} form two equations in the four unknowns $\vec H_{\| 1}$, $\vec E_{\| 1}$, $\vec H_{\| 2}$, and $\vec E_{\| 2}$, in the local coordinate systems

According to (\ref{G}) and (\ref{F}), the self-terms $\dm{G}({\vec \rho _m},{\vec \rho _m})\,\Delta {\ell _m}$ and $\dm{F}({\vec \rho _m},{\vec \rho _m})\,\Delta {\ell _m}$ are singular, and should therefore be treated specifically. This is done in Appendix~\ref{sec:comp_self_terms}, with the result
\begin{align}\label{SG1}
&\dm{G}({\vec \rho _m},{\vec \rho _m})\,\Delta {\ell _m}  =  \nonumber \\
&\begin{pmatrix}
{{{\sin }^2}\psi_m }&{{{\cos \psi_m \sin \psi_m }}/{k}}&0\\
{{{\cos \psi_m \sin \psi_m }}/{k}}&{{{\cos }^2}\psi_m }&0\\
0&0&1
\end{pmatrix} \nonumber \\
&\frac{i}{4}\Delta \ell_m \left\{ 1 + \frac{{2i}}{\pi }\left[\ln \left(\frac{\gamma }{2}k\frac{{\Delta \ell_m }}{2}\right) - 1\right]\right\}  \nonumber \\
+ & \begin{pmatrix}
{{{\cos }^2}\psi_m  - {{\sin }^2}\psi_m }&{ - 2{{\cos \psi_m \sin \psi_m }}/{k}}&0\\
{ - 2{{\cos \psi_m \sin \psi_m }}/{k}}&{{{\sin }^2}\psi_m  - {{\cos }^2}\psi_m }&0\\
0&0&0
\end{pmatrix} \nonumber \\
&\left(\frac{{i\Delta \ell_m }}{8} - \frac{2}{{\pi {k^2}\Delta \ell_m }}\right)
\end{align}
and
\begin{equation}\label{SF1}
\dm{F}({\vec \rho _m},{\vec \rho _m})\,\Delta {\ell _m}  =
\begin{pmatrix}
0&0&-{\cos \psi_m } \\
0&0&-{\sin \psi_m } \\
{\cos \psi_m } & {\sin \psi_m } &0
\end{pmatrix} \frac{\pm 1}{2}
\end{equation}
where, as shown in Fig.~\ref{fig:fields_dist_zero}, $\psi$ is the angle between $-\hat{t}$ and $\hat{x}$ with values between -$\pi$ and $\pi$, $\gamma$ is the Euler constant ($\gamma\simeq$ 1.78107) and the $\pm$ signs in~\eqref{SF1} are used for the exterior region (1) and interior region (2), respectively.

\subsection{Combined GSTC-IE System}

Equations~\eqref{GSTC2} and~\eqref{MM12} form a system of four equations in the four unknowns $\vec E_{\| 1}$, $\vec H_{\| 1}$, $\vec E_{\| 2}$ and $\vec H_{\| 2}$, and is hence fully determined. In order to make it computationally more convenient, we recast this system in a matrix form. For this purpose, we write the tangential fields as $2N\times1$ block vectors including the values at all the discretization points $\vec \rho_m$. For instance, we have
\begin{equation}
{{\bf{E}}_1} = \left( \begin{array}{*{20}{c}}
{\vec E_{\| 1}({\vec \rho_1})}\\
{\vec E_{\| 1}({\vec \rho_2})}\\
{\vec E_{\| 1}({\vec \rho_3})}\\
\vdots \\
{\vec E_{\| 1}({\vec \rho_m})}\\
 \vdots \\
{\vec E_{\| 1}({\vec \rho_N})}
\end{array} \right)
\end{equation}
where $\vec E_{\| 1}({\vec \rho_m})$ is the tangential electric field in region 1 observed at $\vec \rho_m$ and expressed in the local coordinate system. The other tangential vectors, ${\bf{H}}_1$, ${\bf{E}}_2$ and ${\bf{H}}_2,$ are defined similarly. Moreover, the Green function $\dm{G}(\vec \rho _m,\vec \rho _n)$ and its curl $\dm{F}(\vec \rho _m,\vec \rho _n)$ at the source points $\vec \rho _n$ and observation points $\vec \rho _m$ form the $2N\times2N$ matrices $\bf{G}$ and $\bf{F}$, respectively, given by
\begin{subequations}
\begin{equation}
\bold{G} = [\dya{T}^\text{t}({\vec \rho _m})\cdot\dm{G}(\vec \rho _m,\vec \rho _n)\cdot\dya{T}({\vec \rho _n}) \Delta \ell_n]
\end{equation}
and
\begin{equation}
\bold{F}  = [\dya{T}^\text{t}({\vec \rho _m})\cdot\dm{F}(\vec \rho _m,\vec \rho _n)\cdot\dya{T}({\vec \rho _n}) \Delta \ell_n]
\end{equation}
\end{subequations}
in the local coordinate system.

Defining the $2N\times2N$ block diagonal (BDiag) matrices
\begin{subequations}
\begin{equation}
\boldsymbol{\chi} = \text{BDiag}[\dya{\chi}_{\|}({\vec \rho _m})]
\end{equation}
and
\begin{equation}
\boldsymbol{\gamma} = \text{BDiag}[\dya{\text{N}}]
\end{equation}
\end{subequations}
finally allows us to express the GSTCs~\eqref{GSTC2} and the IEs~\eqref{MM12} in the four-equation matrix system
\begin{subequations}\label{Fin1}
\begin{equation}
{{\bf{E}}_1} - {{\bf{E}}_{\text{i}}} = i{k_1}{\eta_1}\,{{\bf{G}}_1}{\boldsymbol{\gamma}}\,{{\bf{H}}_1} + {{\bf{F}}_1}{\boldsymbol{\gamma}}\,{{\bf{E}}_1},
\end{equation}
\begin{equation}
{{\bf{E}}_2} =  - i{k_2}{\eta_2}\,{{\bf{G}}_2}{\boldsymbol{\gamma}}\,{{\bf{H}}_2} - {{\bf{F}}_2}{\boldsymbol{\gamma}}\,{{\bf{E}}_2},\,\,\,\,\,\,\,\,
\end{equation}
\begin{align}
&{\boldsymbol{\gamma}}({{\bf{E}}_1} - {{\bf{E}}_2}) = \nonumber \\
&\frac{{i\omega {\mu _0}}}{{2{\eta_0}}}{\boldsymbol{\chi}}_\text{me}  ({{\bf{E}}_1} + {{\bf{E}}_2}) + \frac{{i\omega {\mu _0}}}{2}{\boldsymbol{\chi}}_\text{mm} ({{\bf{H}}_1} + {{\bf{H}}_2}),
\end{align}
\begin{align}
&{\boldsymbol{\gamma}}({{\bf{H}}_1} - {{\bf{H}}_2}) = \nonumber \\
& - \frac{{i\omega {\varepsilon _0}}}{2}{\boldsymbol{\chi}}_\text{ee}  ({{\bf{E}}_1} + {{\bf{E}}_2}) - \frac{{i\omega }}{{2c}}{\boldsymbol{\chi}}_\text{em}  ({{\bf{H}}_1} + {{\bf{H}}_2}),
\end{align}
\end{subequations}
whose solutions are the tangential electric field set ${\bf{E}}_1$ and ${\bf{E}}_2$ and tangential magnetic field set ${\bf{H}}_1$ and ${\bf{H}}_2$.

If the source is inside region~2, then the first two equations of (\ref{Fin1}) are replaced by
\begin{subequations}\label{Fin2}
\begin{equation}
{{\bf{E}}_1} = i{k_1}{\eta_1}\,{{\bf{G}}_1}{\boldsymbol{\gamma}}\,{{\bf{H}}_1} + {{\bf{F}}_1}{\boldsymbol{\gamma}}\,{{\bf{E}}_1},\,\,\,\,\,\,\,\,\,\,\,\,\,\,\,\,\,\,
\end{equation}
\begin{equation}
{{\bf{E}}_2} - {{\bf{E}}_\text{i}} =  - i{k_2}{\eta_2}\,{{\bf{G}}_2}{\boldsymbol{\gamma}}\,{{\bf{H}}_2} - {{\bf{F}}_2}{\boldsymbol{\gamma}}\,{{\bf{E}}_2},
\end{equation}
\end{subequations}
while the last two are left unchanged. Once the tangential fields have been obtained from (\ref{Fin1}) or (\ref{Fin2}), the fields everywhere, outside or inside of the cavity, are computed using~\eqref{eq:Hp_IE}.

\subsection{Numerical Validation}\label{sec:num_val}
\begin{figure}[t!]
\centerline{\includegraphics[width=1.9 in, height = 1.7 in]{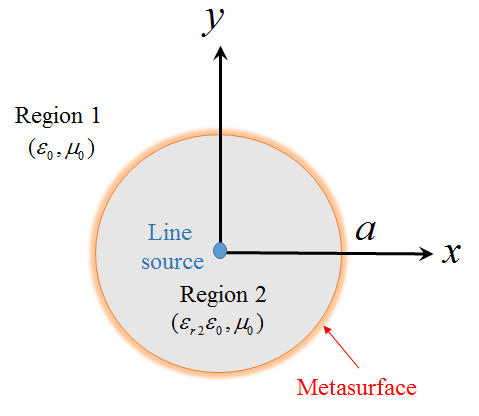}}
\caption{Circular dielectric cylinder of radius $a$ with relative permittivity $\epsilon_{\text{r}2}$ coated by a metasurface and excited by a line source placed at its center.\label{cir}}
\end{figure}

As an initial numerical validation, we compare here the numerical results of Eq.~\eqref{Fin1} and~\eqref{Fin2} with their analytical counterparts for the simple problem of a circular dielectric cylinder excited by a line source placed at its center, as shown in Fig.~\ref{cir}) without and with metasurface coating.

In the case of metasurface coating, we consider the following simple monoisotropic uniform lossy magnetic susceptibility
\begin{equation}\label{eq:bench_Xmm}
\overline{ \overline {\chi}} _{\text{\,mm}}=
\chi _{\text{mm}}^{tt} = 2i/{k_0},\quad
\overline{ \overline {\chi}} _{\text{\,ee}}
=\overline{ \overline {\chi}} _{\text{\,em}}
=\overline{ \overline {\chi}} _{\text{\,me}}=0.
\end{equation}

This problem admits an analytical solution, which may be found by writing the $z$-component of the electric fields ($E_z$ or TM polarization) in regions 1 and 2 as ${E_{1z}}= -A {k_1}{\eta_1}H_0^{(1)}({k_1}r)/4$ and ${E_{2z}}=- {k_2}{\eta_2}H_0^{(1)}({k_2}r)/4+ B{J_0}({k_2}r)$, respectively, and finding their coefficients $A$ and $B$ from the GSTCs~\eqref{GSTC} with~\eqref{MP}, as derived in  Appendix~\ref{sec:comp_coeff}, with solution given by Eqs.~\eqref{AB}.

\begin{figure}[t!]
\begin{center}
  \subfigure[] {
  \includegraphics[width=2.3 in, height = 1.8 in]{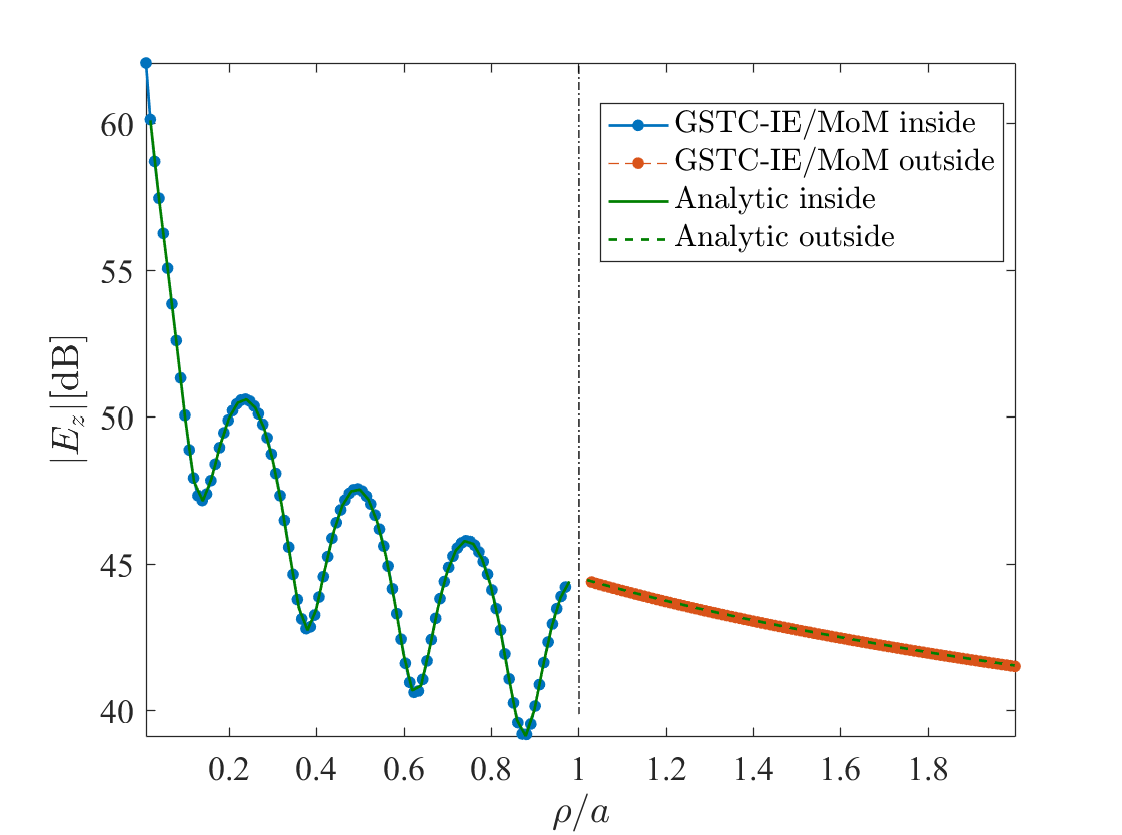}
  }
  \subfigure[] {
  \includegraphics[width=2.3 in, height = 1.8 in]{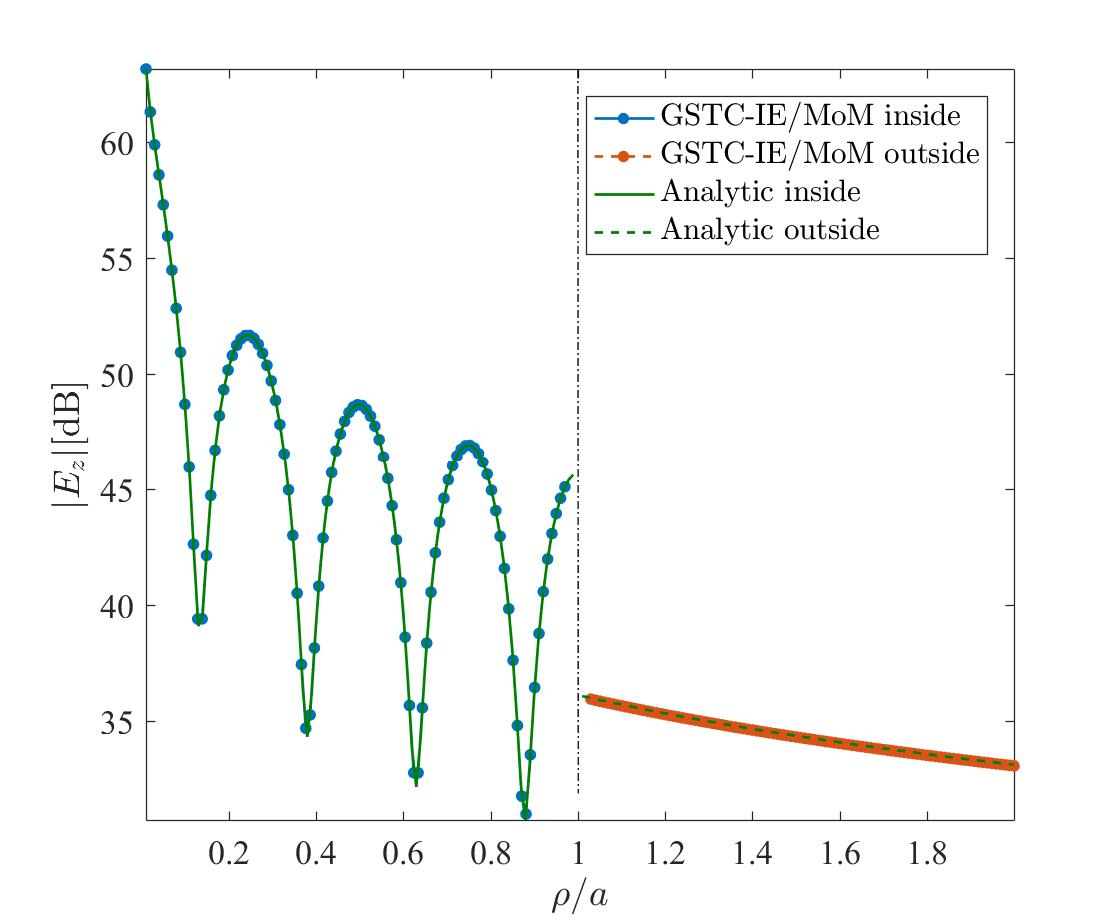}
  }
\caption{Electric field inside ($\rho<a$) and outside ($\rho>a$) of the dielectric cylinder of Fig.~\ref{cir}, (a)~without coating and (b)~coated by the metasurface with the only non-zero tensor component $\chi_{\text{mm}}^{tt}=2i/{k_0}$, for the parameters $f=\omega/(2\pi)=0.3$~GHz, $a=1$, $\varepsilon_{\text{r}2}=4$ and $\varepsilon_{\text{r}1}=1$.
\label{verif}}
\end{center}
\end{figure}

Figures~\ref{verif} plot the GSTC-IE/MoM and analytical results for the line source exciting (a)~the bare dielectric and (b)~that cylinder coated by the metasurface only nonzero susceptibility component $\chi _\text{mm}^{tt}$. Excellent agreement with the analytical solution is observed.
\begin{figure}[h]
\begin{center}
  \subfigure[] {
  \includegraphics[width=2.5 in, height = 2 in]{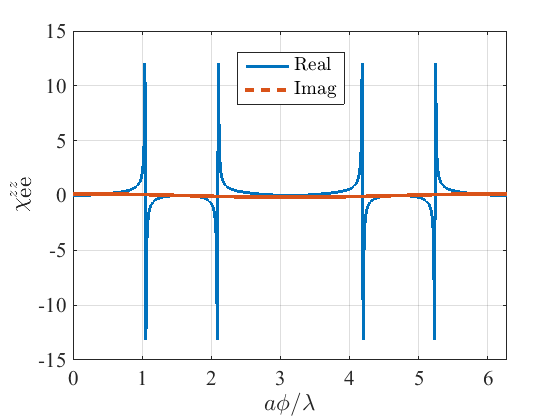}
  }
  \subfigure[] {
  \includegraphics[width=2.5 in, height = 2 in]{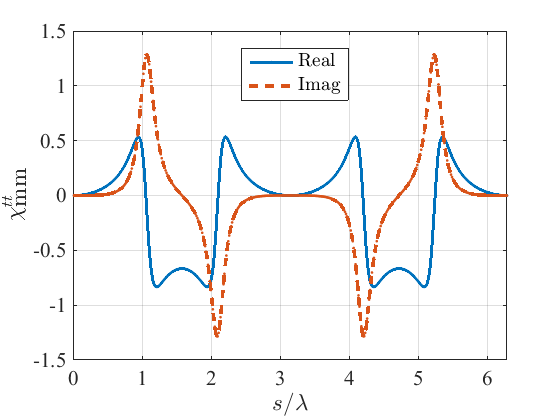}
  }
\caption{Susceptibilities (a)~$\chi _{\text{ee}}^{zz}(\phi)$ and (b)~$\chi _{\text{mm}}^{tt}(\phi)=\chi _{\text{mm}}^{\phi\phi}(\phi)$ along the boundary for the circular cloaking metasurface (Application~A.1) with $+x$-propagating plane wave incidence, and for $a=\lambda$ (cylinder radius), $\epsilon_{\text{r}1}=1$, and $\epsilon_{\text{r}2}=4$.
\label{xcir}}
\end{center}
\end{figure}
\section{Applications}\label{sec:sim}
This section applies the GSTC-IE/MoM technique presented in the previous section to two applications: A)~cloaking and B)~illusion. The two applications further validate the technique, since their results are compared against specified synthesized susceptibilities. All the cases demonstrate the potential of curved metasurfaces. All the numerical results shown will correspond to $E_z$ (or TM) polarization.

\subsection{Cloaking Metasurface}\label{CM}
The first application concerns cloaking metasurfaces, first of circular shape, that would represent the simple type of post, and second of rhombic shape, that may represent for instant struts in parabolic antennas~\cite{KKT}. In both cases, we assume plane wave illumination in the $+x$-direction, corresponding to the incident fields ${E_{1z}} = {e^{ + i{k_1}x}}$ and ${H_{1y}} =  - {e^{ + i{k_1}x}}/{\eta_1}$. In order to cloak the cylinder, the metasurface is synthesized for a total specified field in region~1 everywhere identical to the incident (unperturbed) field, and for a specified field inside the cylinder most reasonably set to ${E_{2z}} = {e^{ + i{k_2}x}}$ (and ${H_{2y}} =  - {e^{ + i{k_2}x}}/{\eta_2}$). Inserting these specifications into~\eqref{GSTC} yields
\begin{equation}\label{sus}
\chi _{\text{ee}}^{zz} = \frac{1}{{i\omega {\varepsilon _0}}}\frac{H_{2t} - H_{1t}} {E_{z,\, \text{av}}}, \quad
\chi _{\text{mm}}^{tt} = \frac{1}{{i\omega {\mu _0}}} \frac{E_{2z} - E_{1z}} {H_{t,\,\text{av}}},
\end{equation}
where ${H_{1t}}$ and ${H_{2t}}$ represent the tangential magnetic fields in regions 1 and 2, respectively.
\begin{figure}[t!]
\begin{center}
  \subfigure[] {
  \includegraphics[width=2.65 in, height = 2 in]{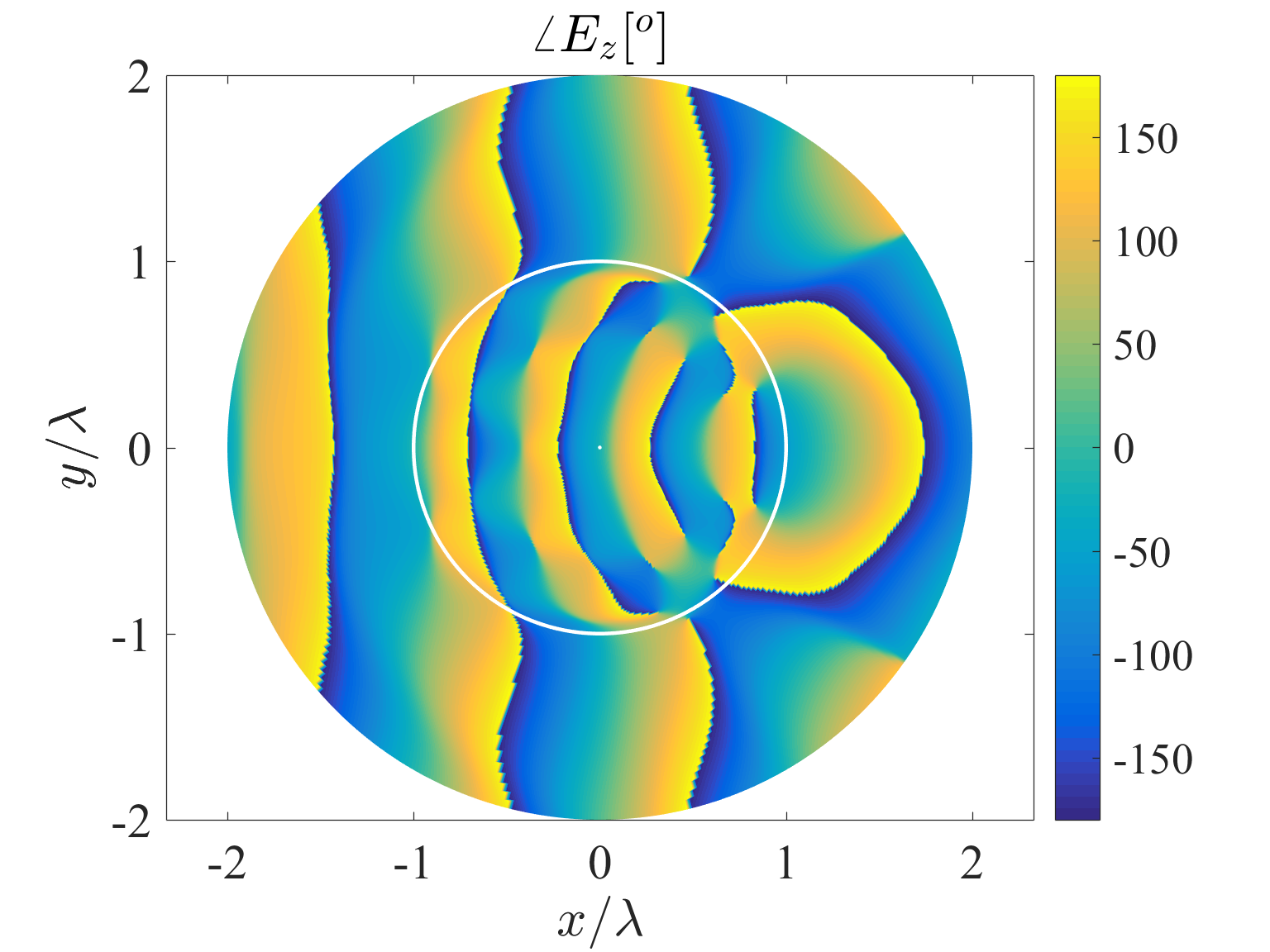}
  }
  \subfigure[] {
  \includegraphics[width=2.65 in, height = 2 in]{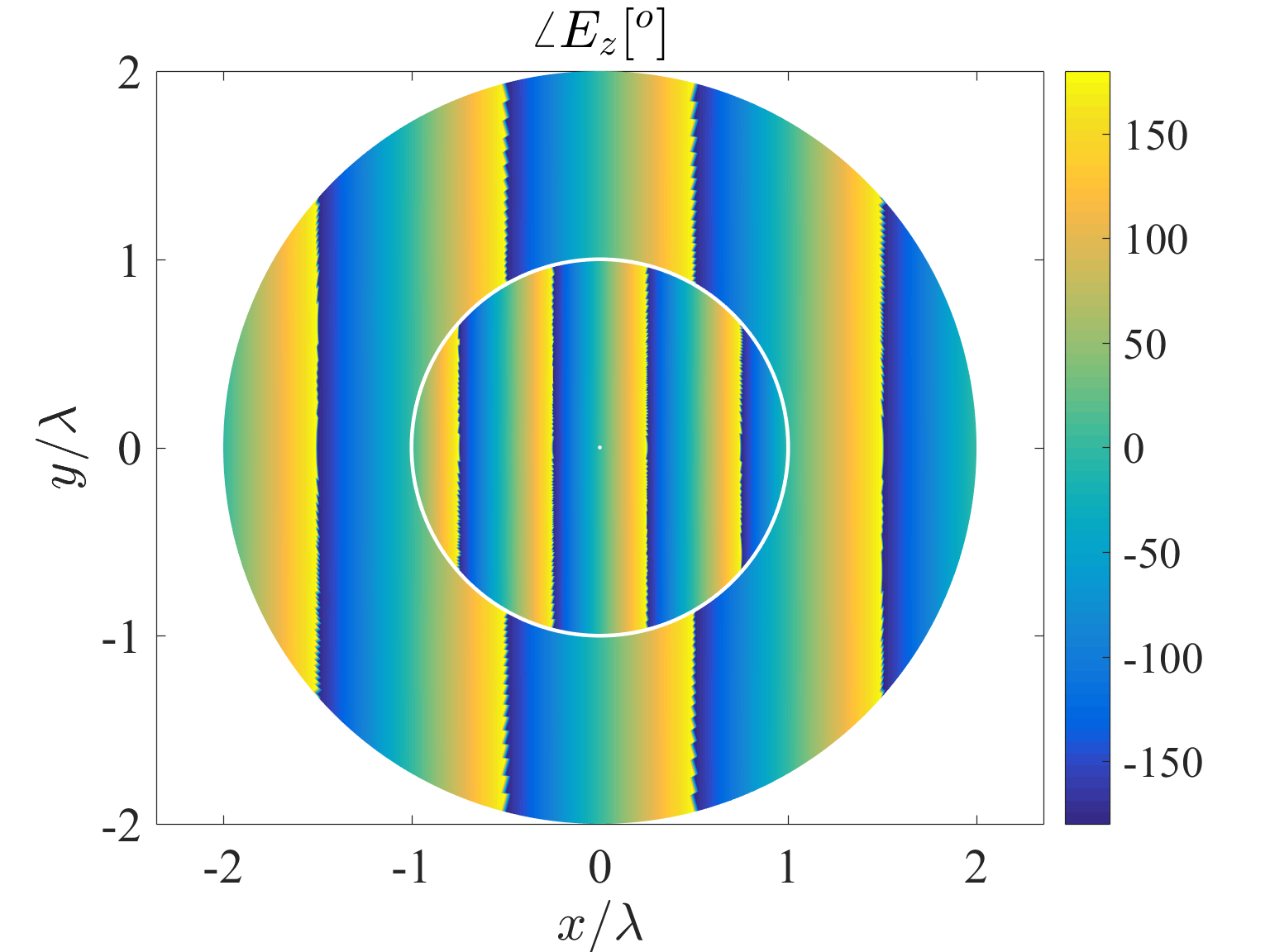}
  }
\caption{Phase distribution of the total electric field in the circular cylindrical cloaking metasurface corresponding to Fig.~\ref{xcir} (a)~without and (b)~with the metasurface. Uniform $\phi$ discretization into $N=250$ arcs, corresponding to $\Delta \ell_n/\lambda_2\approx1/20$.\label{Ecir}}
\end{center}
\end{figure}
\begin{figure}[h]
\begin{center}
\includegraphics[width=2.65 in, height = 2 in]{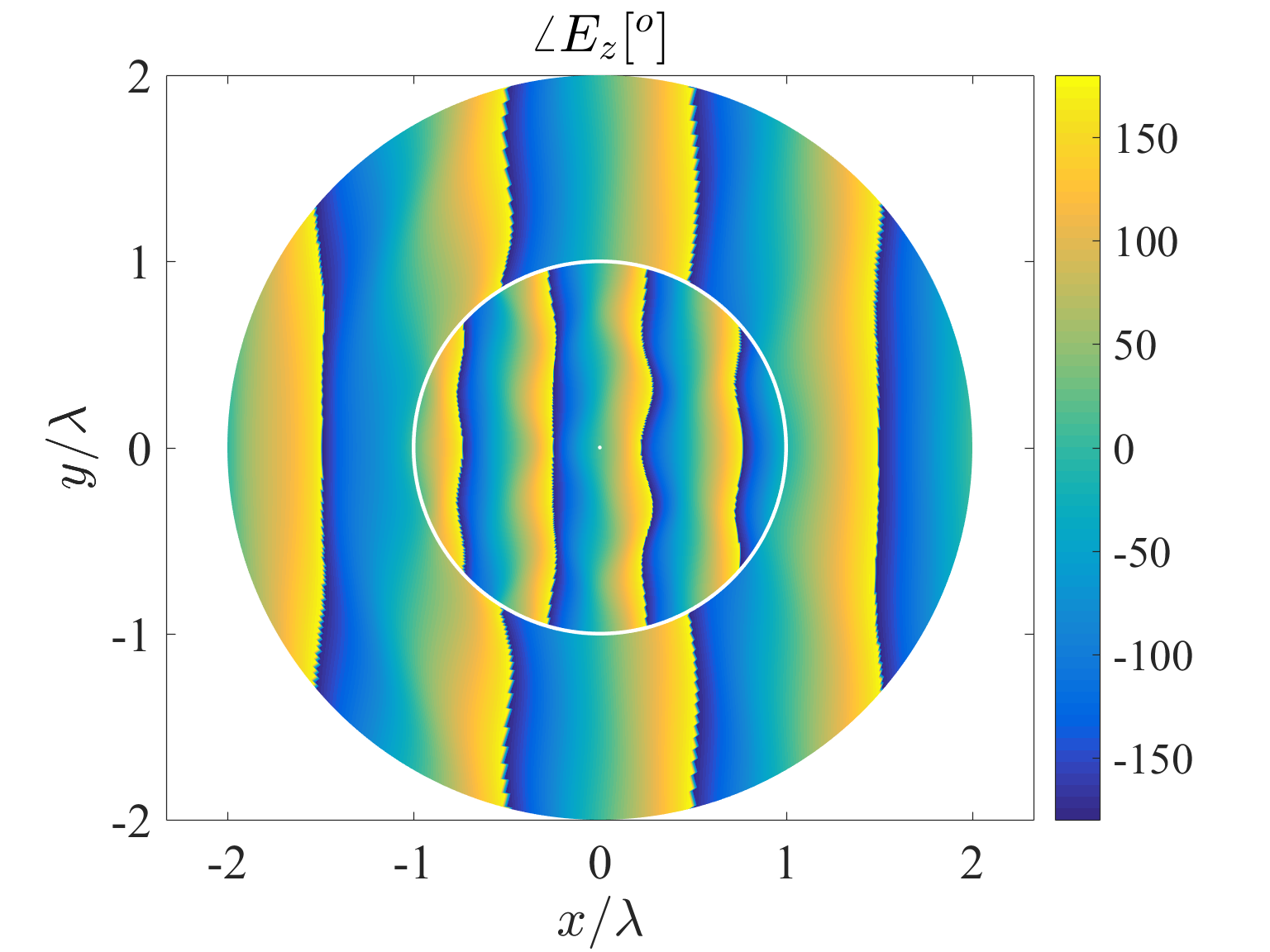}
\caption{Same as Fig.~\ref{Ecir}(b), except the imaginary part of the susceptibility in Fig.~\ref{xcir}(b) has been set to zero at the location where it is negative.\label{PEcir}}
\end{center}
\end{figure}
\subsubsection{Circular Shape}
Figure~\ref{xcir} plots the synthesized susceptibilities~\eqref{sus} for the circular shape (same geometry as in Fig.~\ref{cir}) versus the electrical circular distance from $\phi=0$, i.e. $a\phi/\lambda$. Inserting these susceptibilities into~\eqref{Fin1} provides the fields plotted in Figs.~\ref{Ecir}(a) and~\ref{Ecir}(b) for the cases without and with metasurface, exhibiting the expected scattering and unperturbed phase fronts, respectively.

\begin{figure}[h]
\centerline{\includegraphics[width=2.2 in, height = 1.6 in]{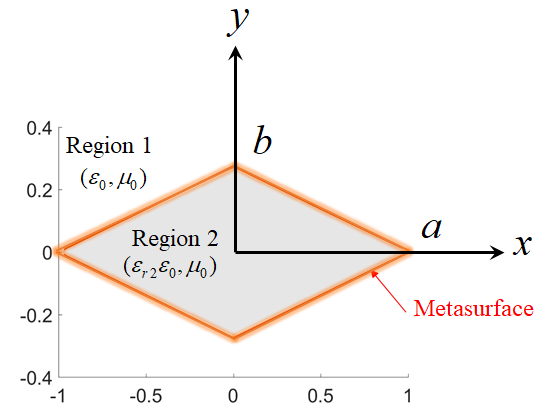}}
\caption{Rhombic cloaking metasurface geometry (Application A.2).\label{rhom}}
\end{figure}
\begin{figure}[h]
\begin{center}
  \subfigure[] {
  \includegraphics[width=2.5 in, height = 2 in]{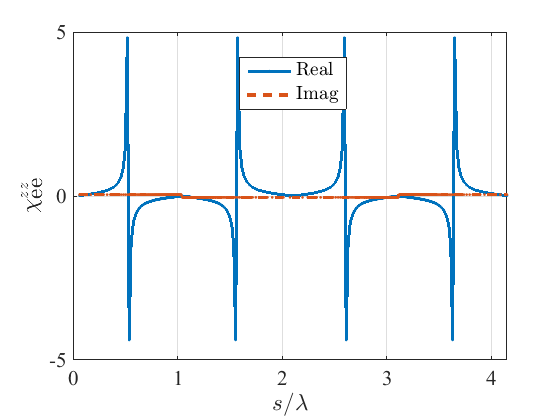}
  }
  \subfigure[] {
  \includegraphics[width=2.5 in, height = 2 in]{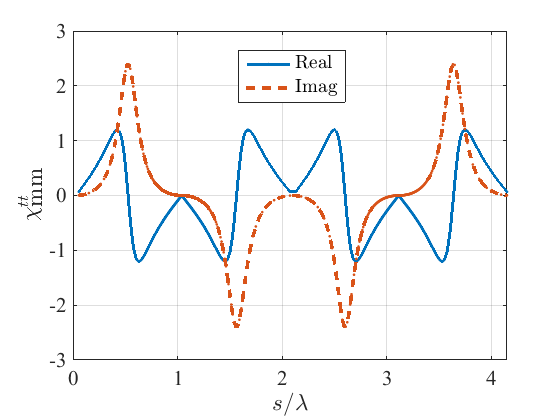}
  }
\caption{Susceptibilities (a)~$\chi _{\text{ee}}^{zz}(s)$ and (b)~$\chi _{\text{mm}}^{tt}(s)$ along the rhombic boundary (Fig.~\ref{rhom}) for $+x$-propagating plane wave incidence, major and minor axes $2a/\lambda_2=4$ and $2b/\lambda_2=1.1$, and $\epsilon_{\text{r}1}=1$ and $\epsilon_{\text{r}2}=4$.\label{Xrhom}}
\end{center}
\end{figure}
\begin{figure}[h]
\begin{center}
  \subfigure[] {
  \includegraphics[width=2.65 in, height = 2 in]{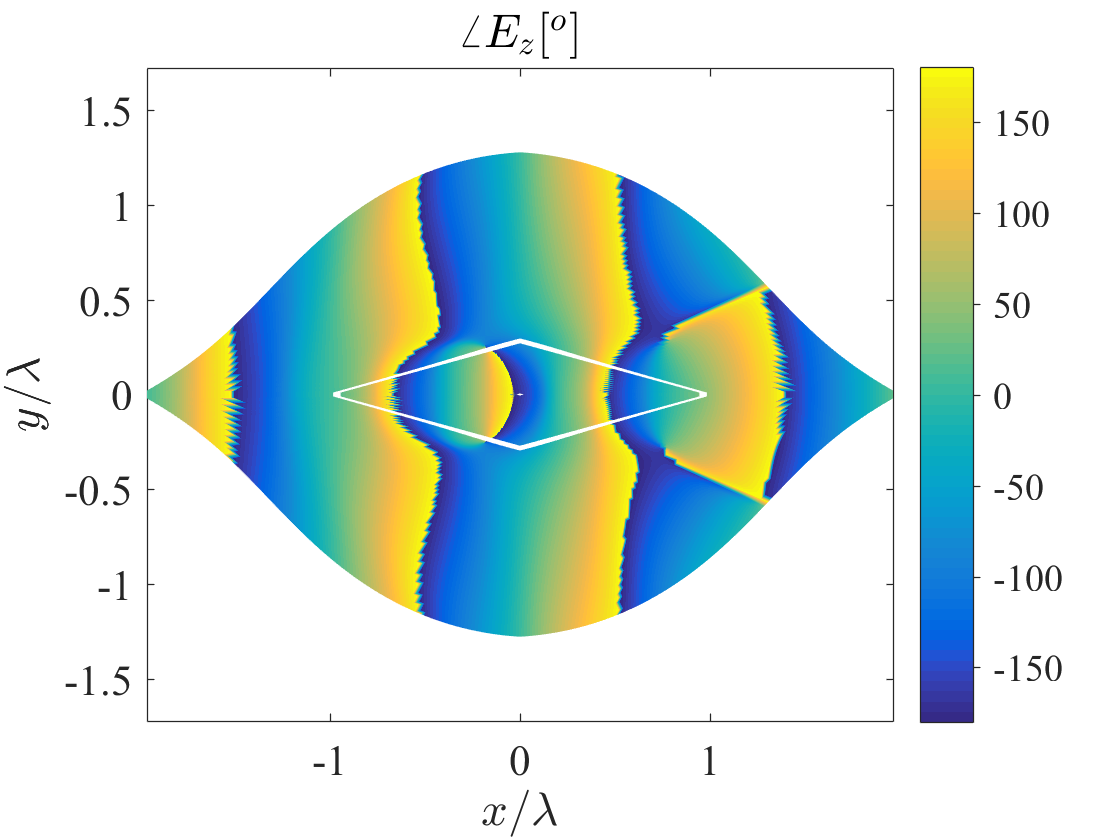}
  }
  \subfigure[] {
  \includegraphics[width=2.65 in, height = 2 in]{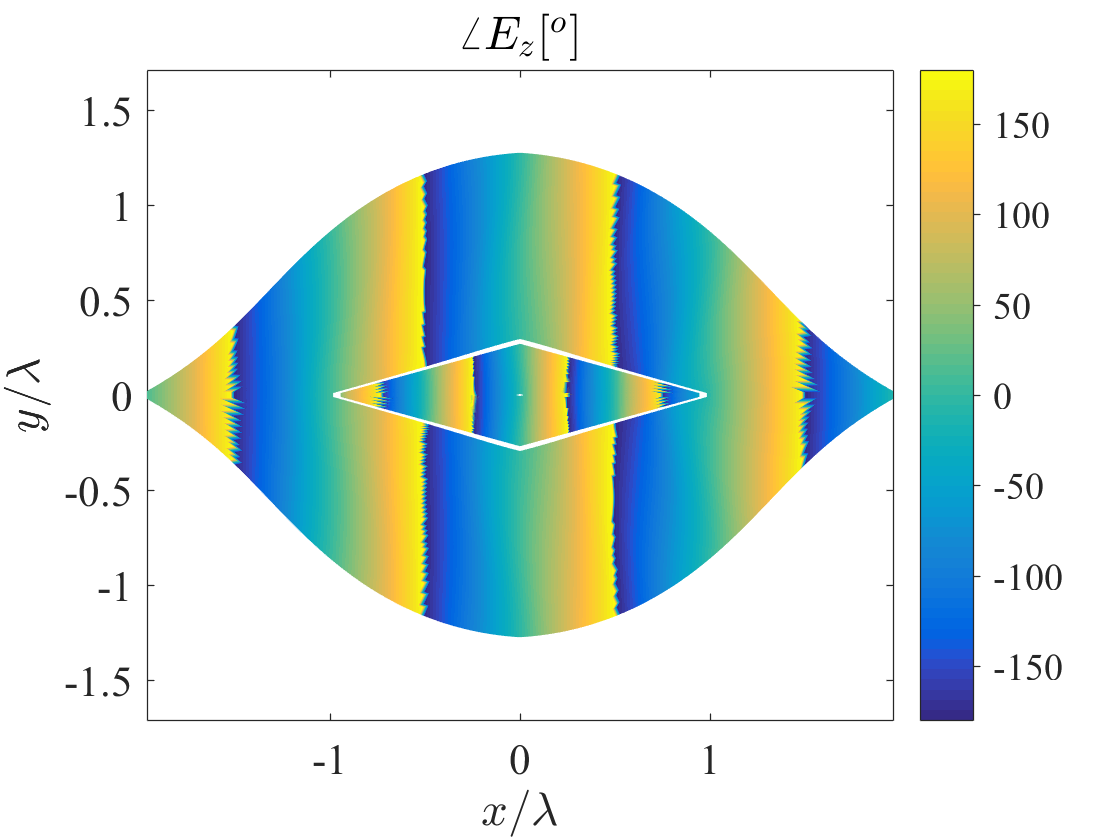}
  }
\caption{Phase distribution of the total electric field in the rhombic cylindrical cloaking metasurface corresponding to Fig.~\ref{Xrhom} (a)~without and (b)~with the metasurface. Uniform rhombic discretization into $N=300$ segments.\label{Erhom}}
\end{center}
\end{figure}
The negative parts of the imaginary susceptibility $\chi_\text{mm}^{tt}$ in Fig.~\ref{xcir}(a) correspond to active ($e^{-i \omega t}$ convention) metasurface sections. They occur in the range $\phi\in[\pi/2,3\pi/2]$, which is the illuminated side of the cylinder, with extrema at $\phi=\pi \pm \pi/3$, where the required wave transformation for cloaking is the most drastic, as seen by comparing the two figures.

As designing active metasurfaces may be challenging, we also consider the closest purely passive design in Fig.~\ref{PEcir}, where the imaginary part of the susceptibility in Fig.~\ref{xcir}(b) is set to zero at the locations where it is negative, with everything else (imaginary part elsewhere and real part everywhere) remaining unchanged. Interestingly, sacrificing the active part of the structure does not induce a drastic penalty in the cloaking response, which reveals that a fairly good two-dimension (2D) cloak may be realized with a purely passive metasurface enclosure!

\subsubsection{Rhombic Shape}

Figure~\ref{rhom} shows the geometry of the rhombic metasurface cavity, and Fig.~\ref{Xrhom} plots the corresponding synthesized susceptibilities~\eqref{sus} versus the electrical rhombic distance from $\phi=0$, $s/\lambda$. Inserting these susceptibilities into~\eqref{Fin1} provides the fields plotted in Fig.~\ref{Erhom}(a) and \ref{Erhom}(b) for the cases without and with metasurface, again exhibiting the expected diffracted and unperturbed phase fronts, respectively.

Similar comments as for the circular shape can be made about the active nature of the illuminated part of the rhombic structure. The scattering response of the passive rhombic structure obtained by setting to zero the imaginary (active) part of the susceptibility in Fig.~\ref{Xrhom} at the location where it is negative is interestingly quite close to that of the perfect active rhombic structure.

\subsubsection{Extinction Cross Width Comparison}

We have qualitatively seen that the aforementioned passive approximation of the cloak still provides a fairly good cloaking effect both in the circular and rhombic metasurface cases. We wish here to quantify this by comparing the \emph{extinction cross widths} of the passive cloaks with those of the corresponding uncoated cylinders.

The scattering cross width may be obtained from the scattering cross section of an electrically long three-dimensional (3D) cylinder via the optical theorem~\cite{Is}, as shown in Appendix~\ref{sec:ext_x_width}. The resulting expression is
\begin{equation}
w_{\text{ext}} = \sqrt {8\pi /{k}}\,{\mathop{\rm Im}\nolimits} \{ e^{- i\pi /4} S_{\text{2D}}(\hat{i},\hat{i})\},
\end{equation}
where $r_\text{ff}$ denotes the far-field distance from the scatterer and $S_{2\text{D}}(\hat{i},\hat{i})$ is the 2D scattering amplitude of the structure in the direction and polarization of the incident wave, or forward scattering amplitude~\cite{Is}.

Table~\ref{tab1} compares the extinction cross width of the circular and rhombic structures without coating, with active metasurface coating and with passive metasurface coating. As expected from synthesis, the active metasurface structure perfectly suppresses the extinction cross width ($w_\text{ext}=0$). However, the passive metasurface, as expected from previous qualitative observations, still remarkably reduces the extinction cross width of the uncoated cylinders at the design wavelength ($\lambda=1$~m). Specifically, the reductions are of about 2.5 and 5.3 or 4~dB and 7~dB for the circular and rhombic cloaks, respectively.

\begin{table}[h]
\caption{Extinction cross width $w_\text{ext}$ for the circular and rhombic cylinders (Applications A.1 and A.2).}
\label{tab1}
\setlength{\tabcolsep}{3pt}
\begin{center}
\begin{tabular}{|p{100pt}|p{50pt}|p{50pt}|}
\hline
$w_\text{ext}(\lambda=1)$&circular case&rhombic case
\\
\hline
No coating&
3.3&1.6\\
Active metasurface&
0 & 0\\
Passive metasurface&
1.3 & 0.3\\
\hline
\end{tabular}
\end{center}
\end{table}

\subsection{Illusion Metasurface}

\begin{figure}[h!]
\centerline{\includegraphics[width=1.8 in, height = 1.5 in]{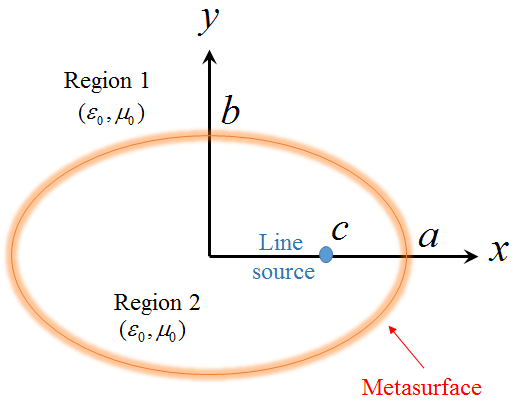}}
\caption{Illusion elliptical metasurface geometry, with line source at the right focus of the ellipse, i.e. at $x=c=\sqrt{a^2-b^2}$ and $y=0$, where $2a$ and $2b$ are the major and minor axes, respectively.\label{ellip} }
\end{figure}

\begin{figure}[h!]
\begin{center}
  \subfigure[] {
  \includegraphics[width=2.5 in, height = 2 in]{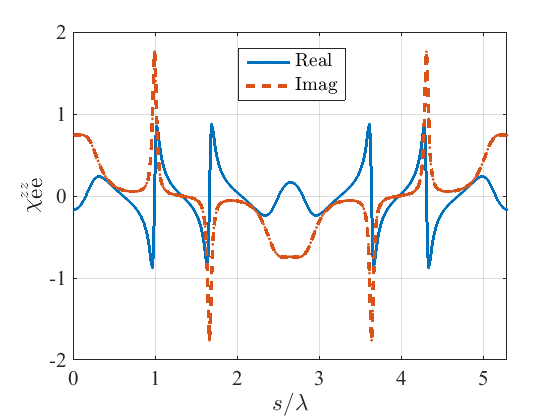}
  }
  \subfigure[] {
  \includegraphics[width=2.5 in, height = 2 in]{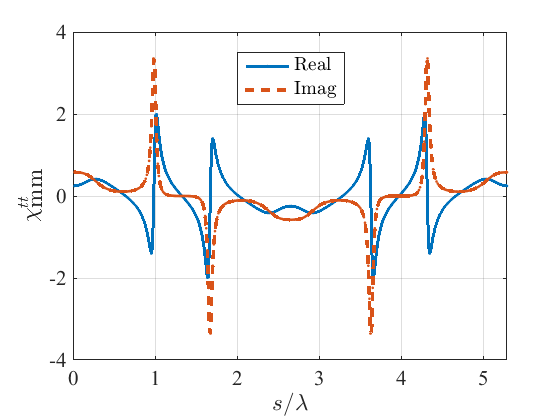}
  }
\caption{Susceptibilities (a)~$\chi _{\text{ee}}^{zz}(s)$ and (b)~$\chi _{\text{mm}}^{tt}(s)$ along the elliptical boundary of the illusion metasurface in Fig.~\ref{ellip}, for major and minor axes $2a/\lambda_=2$ and $2b/\lambda=4/3$, and $\epsilon_{\text{r}1}=\epsilon_{\text{r}2}=1$.\label{Xellip}}
\end{center}
\end{figure}

The next example concerns an elliptical illusion metasurface, with actual source at the right focus and illusion source at the left focus, within the metasurface enclosure. The geometry of the problem is shown in Fig.~\ref{ellip}. The susceptibilities creating this illusion are obtained by inserting into the GSTCs~\eqref{GSTC} the fields for region~1 that would be produced by a source placed at the left focus and the fields for region~2 produced by the actual source at the right focus. Figure~\ref{Xellip} plots the corresponding synthesized susceptibilities.


\begin{figure}[t]
\begin{center}
  \subfigure[] {
  \includegraphics[width=2.65 in, height = 2 in]{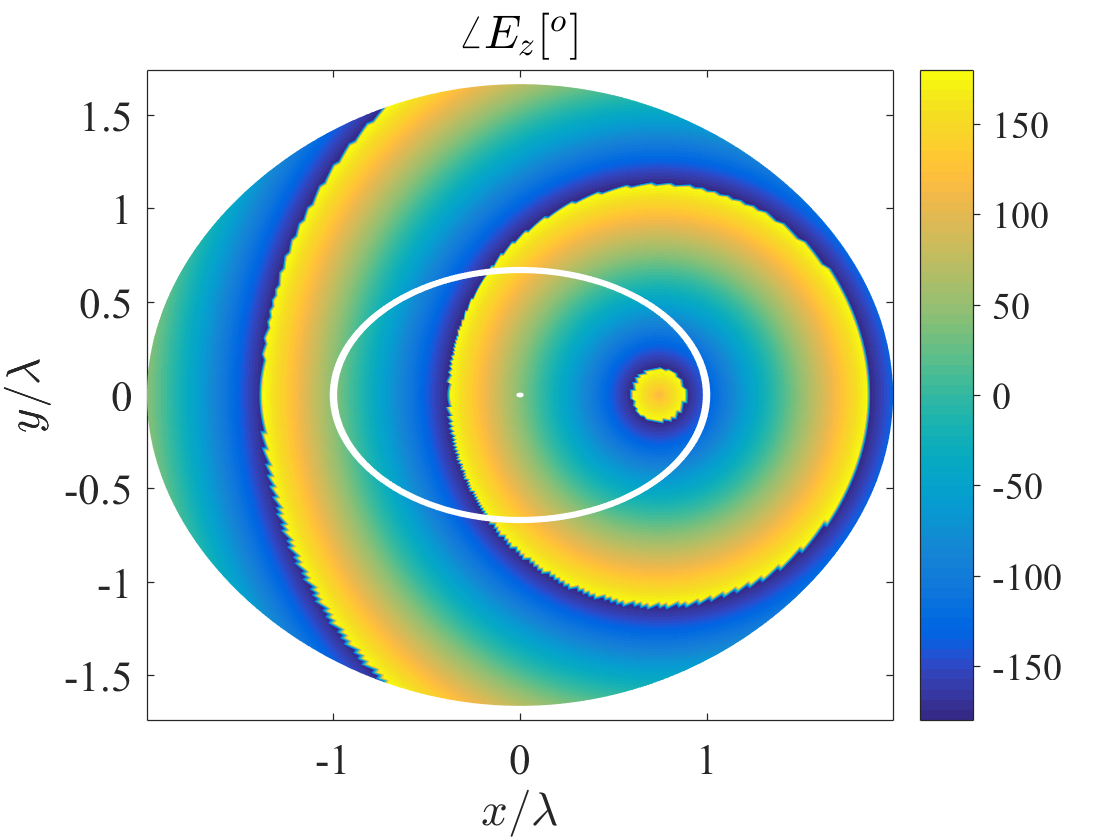}
  }
  \subfigure[] {
  \includegraphics[width=2.65 in, height = 2 in]{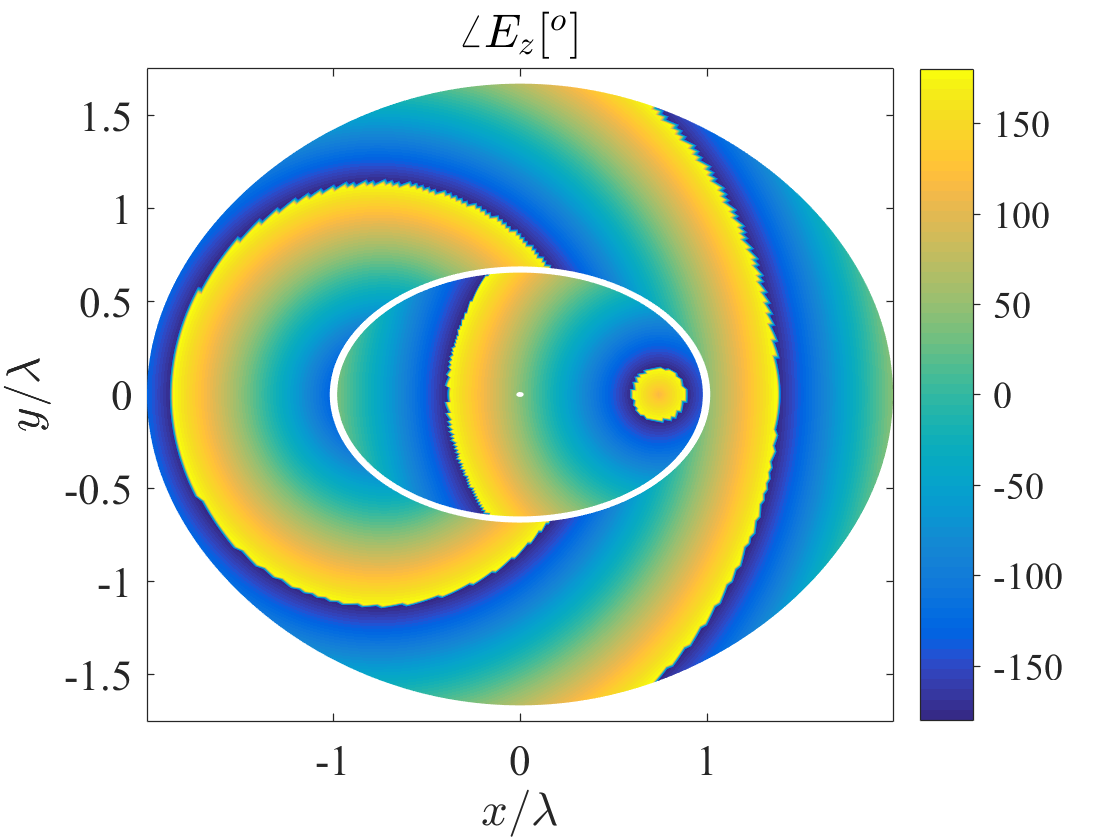}
  }
\caption{Phase distribution of the electric field in the illusion cylindrical metasurface of Fig.~\ref{Xellip} (a)~without and (b)~with the metasurface. \label{Eellip}}
\end{center}
\end{figure}

Figure~\ref{Eellip} shows electric field obtained for this system from~\eqref{Fin1}-\eqref{Fin2}. As specified, the field in region~1 for the illusion metasurface [\ref{Eellip}(b)] is the exact flipped replica of the field produced by a source at the right focus [\ref{Eellip}(a)], corresponding to the illusion of flipped source.

\section{Conclusion}\label{sec:con}

We have presented a technique to compute electromagnetic wave scattering by cylindrical metasurfaces -- forming two-dimensional porous cavities -- of arbitrary cross sections. This technique combines spatial-domain integral equations (IEs) and the Generalized Sheet Transition Conditions (GSTCs) with bianisotropic susceptibility tensors, and leads to a compact and insightful matrix solution. Moreover, we have applied this technique to the problems of cloaking with circular and rhombic shapes and of illusion optics with an elliptic shape. These problems both validate the technique from comparison with specifications and reveal practically significant physical facts. Specifically, metasurfaces are appropriate for both cloaking and illusion optics. Moreover, although such designs are generally active, sacrificing their active features for easier fabrication, leads to performance that is remarkably close to the active design. 

By providing efficient tools for the analysis of arbitrarily curved metasurfaces, this work may lead to the study of a diversity of metasurface cavities, that may be completely closed, porous or open, and feature arbitrary complex shapes. In addition, the physical performance of the metasurface structure demonstrated may stimulate further practical developments in the philosophy of replacing bulk metamaterial by their metasurface counterparts, at least for cloaking and illusion optics.

\appendices
\numberwithin{figure}{section}
\numberwithin{equation}{section}
\section{Derivation of the Self Terms in~\eqref{MM12}~}\label{sec:comp_self_terms}

This section derives analytical expressions for the self terms $\dm{G}({\vec \rho _m},{\vec \rho _m})\Delta {\ell _m}$ and $\dm{F}({\vec \rho _m},{\vec \rho _m})\Delta {\ell _m}$ in~\eqref{MM12}, based on the Green functions~\eqref{G} and~\eqref{F}. For this purpose, consider Fig.~\ref{fig:self_terms}, where the source point is placed on the metasurface segment, while the observation point is initially placed at a small distance $\delta$ from it to avoid singularity.

\subsection{Metasurface Segment Parallel to the $x$ or $y$ Directions}
\begin{figure}[t]\label{fig:self_terms}
\begin{center}
  \subfigure[] {
  \includegraphics[width=2 in, height = 1.05 in]{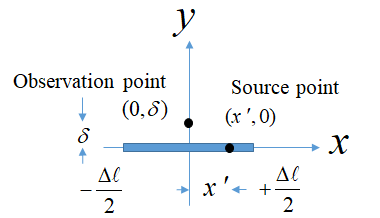}
  }
  \subfigure[] {\label{fig:fields_dist_zero}
  \includegraphics[width=1.2 in, height = 1.05 in]{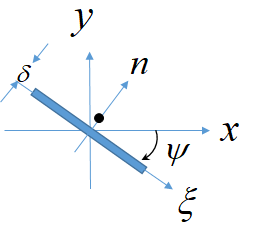}
  }
\caption{Source and observation point positions with respect to a generic metasurface segment of length $\Delta\ell$, with $\delta\rightarrow 0$, for the determination of the self-terms in~\eqref{MM12}. (a)~Segment parallel to the $x$ direction. (b)~Arbitrarily oriented segment, rotated by the angle $\psi$ (here negative).\label{A12}}
\end{center}
\end{figure}

Let us first deal with the simple case shown in Fig.~\ref{A12}(a), where the metasurface segment is parallel to the $x$ direction. We have here $\Delta x = x - x' =  - x'$ and $\Delta y = y - y' = \delta$, and hence $\Delta \rho  = \sqrt{{{x'}^2} + {\delta^2}}$, where $\delta\rightarrow 0$. The argument of the Hankel functions in~\eqref{G} and~\eqref{F}, $s=k\Delta\rho=k\sqrt{{{x'}^2}+{\delta^2}}$ is then infinitesimally small, and one may therefore use the small-argument asymptotic approximations
\begin{subequations}\label{eq:Hankel_asymp}
\begin{equation}\label{eq:Hankel0_asymp}
{\lim _{s \to 0}}H_0^{(1)}(s) = 1 + \frac{{2i}}{\pi }\ln \left(\dfrac{\gamma s}{2}\right),
\end{equation}
\begin{equation}\label{eq:Hankel1_asymp}
{\lim _{s \to 0}}H_1^{(1)}(s) = \frac{s}{2} - \frac{{2i}}{{\pi s}},
\end{equation}
\end{subequations}
where $\gamma$ is the Euler constant ($\gamma \simeq$ 1.78107). Substituting~\eqref{eq:Hankel0_asymp} and~\eqref{eq:Hankel1_asymp} into~\eqref{G} and~\eqref{F}, respectively, and further substituting the result into~\eqref{eq:Hp_IE} leads to integrals of the type
\begin{equation}\label{eq:lims}
\begin{split}
&\lim _{\delta \to 0} \Bigg[ \frac{i}{4} \int\limits_{- \frac{{\Delta \ell }}{2}}^{\frac{{\Delta \ell }}{2}}
\frac{\delta ^2}{{x'^2}+\delta ^2} \left[ 1+\frac{2i}{\pi}\ln\left(\frac{\gamma}{2}k\sqrt{{x'^2}+\delta^2} \right) \right ]\,\,dx' \\
&\quad+ \frac{i}{4k} \int\limits_{- \frac{{\Delta \ell }}{2}}^{\frac{{\Delta \ell }}{2}}\frac{x'^2-\delta^2}{\sqrt{x'^2+\delta^2}^3} \left( \frac{k\sqrt{x'^2+\delta^2}}{2}-\frac{2i}{\pi k \sqrt{x'^2+\delta^2}} \right)\,\,dx' \Bigg].
\end{split}
\end{equation}
for both $\dm{G}$ and $\dm{F}$. It may be easily verified that the first integral in~\eqref{eq:lims} is $\mathcal{O}(\delta)$ (proportional to $\delta$) and hence zero from the limit. The second integral has two terms. It may be easily found that the first term tends to ${i\Delta \ell }/{8}$ while the second term tends to $-{2}/{\pi {k^2}\Delta \ell}$. Similar calculations are performed for the other terms in~\eqref{G} and~\eqref{F}. It is noted that the integrals of the off-diagonal elements of $\dm{G}$ are zero since they are odd functions of $x'$. After some algebraic manipulations, one finally obtains
\begin{align}\label{eq:limG}
{\lim _{\delta  \to 0}}\int\limits_{- \frac{{\Delta \ell }}{2}}^{\frac{{\Delta \ell }}{2}} & {\dm{G}\,\,dx'}  =  \nonumber \\
& \left( {\begin{array}{*{20}{c}}
0&0&0\\
0&1&0\\
0&0&1
\end{array}} \right) \frac{i}{4}\Delta \ell \left[ 1 + \frac{{2i}}{\pi } \left[ \ln \left( \frac{\gamma }{2} k \frac{{\Delta \ell }}{2} \right) - 1 \right] \right] \nonumber \\
+& \left( {\begin{array}{*{20}{c}}
1&0&0\\
0&{ - 1}&0\\
0&0&0
\end{array}} \right) \,\, \left( \frac{{i\Delta \ell }}{8} - \frac{2}{{\pi {k^2}\Delta \ell }} \right)
\end{align}
and
\begin{equation}\label{eq:limF}
{\lim _{\delta  \to 0}}\int\limits_{- \frac{{\Delta \ell }}{2}}^{\frac{{\Delta \ell }}{2}} {\dm{F}\,\,dx'}  =
\left( {\begin{array}{*{20}{c}}
0&0&-1\\
0&0&0\\
1&0&0
\end{array}} \right) \frac{ \text{sgn} (\delta)}{2},
\end{equation}
where $\text{sgn}(\cdot)$ is the sign function.

\subsection{Metasurface Segment with Arbitrary Orientation}
If the metasurface is rotated with respect to the coordinate system $(x,y)$, we use the new coordinates $(\xi,n)$, as shown in Fig.~\ref{A12}(b), where $\hat \xi  =  - \hat t$ (Fig.~1), and where the angle $\psi$ between $\hat \xi$ and $\hat x$ varies between -$\pi$ and $\pi$. The two coordinate systems are related as
\begin{equation}\label{rot}
\left( {\begin{array}{*{20}{c}}
x\\
y
\end{array}} \right) = \left( {\begin{array}{*{20}{c}}
{\cos \psi }&{ - \sin \psi }\\
{\sin \psi }&{\cos \psi }
\end{array}} \right)\left( {\begin{array}{*{20}{c}}
\xi \\
n
\end{array}} \right),
\end{equation}
which also imply $\Delta x = \,\cos \psi \,\Delta \xi  - \sin \psi \,\Delta n$ and $\Delta y = \,\sin \psi \,\Delta \xi  + \cos \psi \,\Delta n$.  Substituting these expressions for $\Delta x$ and $\Delta y$  into~\eqref{G} and~\eqref{F}, and simplifying, yields
\begin{align}\label{SG}
&{\lim _{\delta  \to 0}}\int\limits_{- \frac{{\Delta \ell }}{2}}^{\frac{{\Delta \ell }}{2}}  {\dm{G}\,\,d\xi'}  =  \nonumber \\
& \left( {\begin{array}{*{20}{c}}
{{{\sin}^2}\psi}&{{{\cos \psi \sin \psi }}/{k}}&0\\
{{{\cos \psi \sin \psi }}/{k}}&{{{\cos }^2}\psi }&0\\
0&0&1
\end{array}} \right) \nonumber \\
&\frac{i}{4}\Delta \ell \left[ 1 + \frac{{2i}}{\pi } \left[ \ln \left( \frac{\gamma }{2} k \frac{\Delta \ell }{2} \right) - 1 \right] \right]  \nonumber \\
+ & \left( {\begin{array}{*{20}{c}}
{{{\cos }^2}\psi  - {{\sin }^2}\psi }&{ - 2{{\cos \psi \sin \psi }}/{k}}&0\\
{ - 2{{\cos \psi \sin \psi }}/{k}}&{{{\sin }^2}\psi  - {{\cos }^2}\psi }&0\\
0&0&0
\end{array}} \right) \nonumber \\
& \left( \frac{i\Delta \ell }{8} - \frac{2}{\pi {k^2}\Delta \ell } \right)
\end{align}
and
\begin{equation}\label{SF}
{\lim _{\delta  \to 0}}\int\limits_{- \frac{{\Delta \ell }}{2}}^{\frac{{\Delta \ell }}{2}} {\dm{F}\,\,d\xi'}  =
\left( {\begin{array}{*{20}{c}}
0&0&-{\cos \psi } \\
0&0&-{\sin \psi } \\
{\cos \psi } & {\sin \psi } &0
\end{array}} \right) \frac{\text{sgn}(\delta)}{2},
\end{equation}
which reduce to~\eqref{eq:limG} and~\eqref{eq:limF}, respectively, for $\psi=0$.

\section{Derivation of the Scattered Fields in Sec.~\ref{sec:num_val}}\label{sec:comp_coeff}

This section derives the analytical solution for the fields scattered by the metasurface-coated [susceptibilities~\eqref{eq:bench_Xmm}] circular dielectric cylinder excited by a centered line source, that is used as a benchmark in Sec.~\ref{sec:num_val}.

Assuming $E_z$ or TM polarization, the electric fields in regions 1 and 2 are respectively given by
\begin{subequations}\label{E12}
\begin{equation}
{E_{1z}}= -A \frac{k_1 \eta_1}{4}H_0^{(1)}({k_1}r),
\end{equation}
\begin{equation}
{E_{2z}}=- \frac {k_2 \eta_2}{4}H_0^{(1)}({k_2}r)+ B{J_0}({k_2}r),
\end{equation}
\end{subequations}
and are associated with the $\phi$ components of the magnetic field

\begin{subequations}\label{H12}
\begin{equation}
{H_{1\phi}}= \frac{i k_1}{4}A H_1^{(1)}({k_1}r),
\end{equation}
\begin{equation}
{H_{2\phi}}= \frac{i k_2}{4} H_1^{(1)}({k_2}r)- \frac{i B}{\eta_2}{J_1}({k_2}r),
\end{equation}
\end{subequations}
where coefficients $A$ and $B$ are found upon inserting these relations into the GSTCs~\eqref{GSTC} with~\eqref{MP}, which yields
\begin{subequations}\label{sGSTC}
\begin{equation}
{E_{1z}}-{E_{2z}}=-i\omega\mu_0\chi _{\text{mm}}^{tt}\left(\frac{{H_{1\phi}}+{H_{2\phi}}}{2}\right),
\end{equation}
\begin{equation}
{H_{1\phi}}-{H_{2\phi}}=0.
\end{equation}
\end{subequations}

Inserting~\eqref{E12} and~\eqref{H12} into~\eqref{sGSTC}, and simplifying, yields then
\begin{subequations}\label{AB}
\begin{equation}
A=\frac{2i\eta_2/k_1\pi a}{C\eta_1{H_0}^{(1)}({k_1}a){J_1}({k_2}a)-\eta_2{H_1}^{(1)}({k_1}a){J_0}({k_2}a)},
\end{equation}
\begin{equation}
B=\frac{\eta_2}{4{J_1}({k_2}a)}\left[ k_2 {H_1}^{(1)}({k_2}a) - A k_1 {H_1}^{(1)}({k_1}a) \right],
\end{equation}
\end{subequations}
where $C=1+\chi _{\text{mm}}^{tt}k_0\eta_0{H_1}^{(1)}({k_1}a)/\eta_1{H_0}^{(1)}({k_1}a)$. The non-coated case naturally corresponds to the particular case \mbox{$\chi _{\text{mm}}^{tt}=0$}.

\section{extinction cross width}\label{sec:ext_x_width}

Section~\ref{sec:formul} solved the general 2D problem of scattering by a metasurface cavity of arbitrary cross section. We wish here to determine the \emph{extinction cross width} -- or 2D counterpart of the extinction cross section -- of such a metasurface cavity in terms of its forward scattering amplitude~\cite{Is}, that is available from solving~\eqref{Fin1}.

In the far-field, the field scattered by the 2D metasurface cavity is
\begin{equation}\label{eq:Es2D}
E_{\text{s2D}} = S_{2\text{D}}e^{i{k}{r_\text{ff}}}/\sqrt{r_{\text{ff}}},
\end{equation}
where $k$ is the free-space wavenumber, $r_\text{ff}$ denotes the (far-field) distance from the scatterer, and $S_{2\text{D}}$ is the 2D scattering amplitude of the structure.

Similarly, the 3D scattered far field is expressed as
\begin{equation}\label{eq:Es3D}
E_\text{s3D}
=S_{\text{3D}}~e^{i{k}{r_\text{ff}}}/r_\text{ff},
\end{equation}
where $S_{\text{3D}}$ is the conventional 3D scattering amplitude.

In the case of an electrically long cylinder, of length $\ell$, where edge diffraction is negligible, the forward scattering amplitudes in~\eqref{eq:Es2D} and~\eqref{eq:Es3D} are related by~\cite{KST}
\begin{equation}\label{S3D}
S_{\text{3D}} = \frac{\ell }{\sqrt \lambda}{e^{- i\pi/4}} S_{\text{2D}},
\end{equation}
where the factor $e^{- i{\pi}/{4}}/\sqrt{\lambda}$ originates in the asymptotic approximation of the 2D scalar Green function (\ref{2DG}) for large arguments, $H_0^{(1)}(kx)\sim\sqrt{2/\pi k x}~e^{ikx-i\pi/4}$.

According to the optical theorem~\cite{Is}, the extinction cross section of an object is related to the 3D forward scattering amplitude as
\begin {equation}\label{sigma_ext}
\sigma _\text{ext} = \frac{{4\pi }}{k}{\mathop{\rm Im}\nolimits} \{S_{3\text{D}}(\hat{i},\hat{i})\},
\end{equation}
where $S_{3\text{D}}(\hat{i},\hat{i})$ is the scattering amplitude with the same direction (here $\phi=0$) and polarization (here $\hat{z}$) as the incident wave, which is also called the forward scattering amplitude~\cite{Is}.

Inserting (\ref{S3D}) in (\ref{sigma_ext}) and dividing by $\ell$ finally yields the sought after extinction cross width
\begin {equation}
{w_\text{ext}} =\sqrt{\frac{8\pi}{k}}\,{\mathop{\rm Im}\nolimits} \{ e^{- i\pi /4} S_\text{2D}(\hat{i},\hat{i})\}.
\end{equation}

\end{document}